\documentclass[prb,twocolumn,showpacs,preprintnumbers,amsmath,amssymb]{revtex4}
\usepackage{graphicx}
\usepackage{dcolumn}
\usepackage{bm}

\begin{document}

\newcommand{\AR}[1]{{\bf B:} #1 {\bf END}}
\newcommand{\COMMENT}[1]{{\bf #1}}

\newcommand{\w}{\omega}
\newcommand{\W}{\Omega}
\newcommand{\e}{\varepsilon}
\newcommand{\s}{\sigma}
\newcommand{\al}{\gamma}
\newcommand{\RE}{\text{Re}}
\newcommand{\IM}{\text{Im}}
\newcommand{\g}{\alpha}
\newcommand{\G}{\Gamma}
\newcommand{\I}{{\cal I}}
\newcommand{\K}{{\cal K}}
\newcommand{\Nf}{N(0)}
\newcommand{\la}{\lambda}
\newcommand{\La}{\Lambda}
\newcommand{\be}{\begin{equation}}
\newcommand{\bea}{\begin{eqnarray}}
\newcommand{\ee}{\end{equation}}
\newcommand{\eea}{\end{eqnarray}}
\newcommand{\bwt}{\begin{widetext}}
\newcommand{\ewt}{\end{widetext}}
\newcommand{\ind}[5]{{_{#1}}\!\!\!\!{^{#2}}#3{_{#4}}\!\!\!\!\!{^{#5}}}
\newcommand{\ovrl}{\overline}
\newcommand{\undl}{\underline}
\newcommand{\p}{\prime}
\newcommand{\up}{\uparrow}
\newcommand{\down}{\downarrow}
\newcommand{\gd}{g_{d}}
\newcommand{\gll}{g_{LL}}
\newcommand{\grr}{g_{RR}}
\newcommand{\glr}{g_{LR}}
\newcommand{\Gpf}{{\hat {\cal G}}}
\newcommand{\Lpf}{{\hat L}}
\newcommand{\GR}{{\cal G}^{R}}
\newcommand{\GA}{{\cal G}^{A}}
\newcommand{\GK}{{\cal G}^{K}}
\newcommand{\GL}{{\cal G}^{<}}
\newcommand{\GG}{{\cal G}^{>}}


\title{Nonequilibrium Transport through a Kondo Dot in a Magnetic Field:\\
       Perturbation Theory}

\author{
J. Paaske
, A. Rosch 
and P. W\"olfle
}

\affiliation{Institut f\"ur Theorie der Kondensierten Materie, 
Universit\"at Karlsruhe, D-76128 Karlsruhe, Germany}

\date{\today}

\begin{abstract}
  Using nonequilibrium perturbation theory, we investigate the
  nonlinear transport through a quantum dot in the Kondo regime in
  the presence of a magnetic field.  We calculate the leading logarithmic
  corrections to the
  local magnetization and the differential conductance, which are
  characteristic of the Kondo effect out of equilibrium.  By solving a
  quantum Boltzmann equation, we determine the nonequilibrium
  magnetization on the dot and show that the application of both a finite bias
  voltage and a magnetic field induces a novel structure of
  logarithmic corrections not present in equilibrium.  These
  corrections lead to more pronounced features in the conductance,
  and their form calls for a modification of the
  perturbative renormalization group.
\end{abstract}


\maketitle

A localized spin coupled to the spins of a conduction electron system
via a Heisenberg exchange interaction is known to give rise to the
Kondo effect, provided the coupling is
antiferromagnetic\cite{Hewson93}.  At sufficiently high temperatures
or energies, well above the characteristic energy scale referred to as
the Kondo temperature, $T_K$, the signature of the Kondo effect is a
logarithmic variation of various observables with temperature (or
other energies).  As first demonstrated by Kondo\cite{Kondo64}, such
logarithmic behavior appears already in perturbation theory to low
order in the exchange coupling.  At low temperatures, however, for $T
\ll T_K$, the local spin is screened by the conduction electron spins,
and the system enters a local Fermi liquid state, characterized by
integer temperature power laws.

The many-body resonance state forming near $T_K$ is comprised by
infinitely many virtual particle hole excitations leading to a peak in
the conduction electron scattering amplitude at the Fermi energy: the
Kondo resonance. In equilibrium systems the Bethe ansatz method allows
to analytically calculate thermodynamic properties\cite{Andrei8082},
and dynamical properties can be determined with the help of Wilson's
numerical renormalization group\cite{Wilson7580}. These methods reveal
the universal nature of the Kondo effect: a single characteristic
scale $T_K$ determines the physics and all physical quantities are
universal functions of, e.g., $T/T_K$. Already in 1970
Anderson\cite{Anderson70} suggested a simple and efficient method,
known as ``poor man's scaling'' or perturbative renormalization group,
to resum the leading logarithmic terms in perturbation theory and to
establish the scaling behavior.  It requires nothing more than low
order perturbation theory, and provides a controlled approximation for
$T >T_K$, i.e. as long as the running exchange couplings remain small.

While the Kondo effect was discovered in metals containing magnetic
impurities in the 1960's (with experimental observations dating back
to the 1930's), and most of the theoretical development took place
before the mid 1980's, it had a revival in the 1990's in the context
of electron transport through quantum dots weakly coupled to leads.
Provided that the dot carries a net spin, a
Kondo resonance develops, thus admitting resonant tunneling of
electrons\cite{Glazman88,Ng88}.  This leads to a removal of the
Coulomb blockade, i.e. an increase of the conductance from small
values up to the quantum limit, as the Kondo resonance develops.  We
shall refer to this type of quantum dot as a {\it Kondo dot}.  It has
been observed that the Kondo effect is quenched by raising the
transport bias voltage V well above $T_K$, i.e. $eV \gg T_K$, and that
the presence of a magnetic field splits the zero-bias conductance peak
into two distinct peaks, located at bias voltages roughly equal to plus
and minus the Zeeman splitting of the spin on the
dot\cite{Goldhaber98,Cronenwett98,Schmid98,Nygaard00,vanderWiel00}.
Furthermore, it has proven possible to collapse the data from
different Coulomb blockade valleys onto one universal conductance
curve\cite{Goldhaber98b,vanderWiel00,Nygaard00}, thus establishing the
aforementioned one-parameter scaling out of equilibrium.

The observation of {\it Kondoesque} conductance anomalies, exhibiting
a logarithmic temperature dependence of the zero-bias peak height and
a Zeeman splitting of the peak in finite magnetic field, actually has
a much longer history; it has often been observed in more traditional
tunnel junctions involving tunneling via magnetic impurities. In
metal-insulator-metal junctions the Kondo-anomaly can result from
magnetic impurities present in the insulating metal-oxide barrier,
from surface states at the metal/metal-oxide
interface\cite{Shen68,Nielsen70,Appelbaum72,Wallis74,Bermon78} or even
from unpaired electrons residing on organic radicals in a polymerized
benzene barrier\cite{Magno77}. The same type of conductance anomaly
has been detected in semiconductor-metal (Schottky)
junctions\cite{Wolf70,Wolf75}, in which the neutral shallow donors in
the semiconductor depletion layer provide the spin-$1/2$ moments which
incite the Kondo correlations.

In 1966, Appelbaum\cite{Appelbaum66,Appelbaum67a} and
Anderson\cite{Anderson66} demonstrated that this type of conductance
anomaly can result from so called {\it exchange tunneling} processes,
in which an electron tunnels from one electrode to the other via an
intermediate magnetic impurity orbital and at the same time flips its
spin. This mechanism was shown to lead to a Kondo effect enhancing the
charge transfer across the dot, and it explains in a natural way why
the conductance is peaked at a bias voltage corresponding to the
Zeeman splitting of the impurity moment: the finite bias has to supply
the energy to flip the spin in the presence of a magnetic field.
Appelbaum calculated the tunneling conductance from a simple golden
rule expression, in which the tunneling amplitudes were determined
from third order perturbation theory,
including the leading logarithmic corrections reflecting the Kondo
effect. His result did capture the qualitative features of the
conductance anomaly, but any quantitative agreement with experiment
has been restricted to the case of zero magnetic field.  Wolf and
Losee\cite{Losee69} later suggested an extension of Appelbaum's theory
in which the logarithmic enhancement was cut off by the lifetime
broadening of the Zeeman split spin-levels. Indeed this extra feature
improved the agreement with experiments somewhat, but even with
several parameters, this did not allow for a fit to experiments at
finite magnetic
field\cite{Wolf70,Appelbaum72,Wallis74,Bermon78,Ivezic80}.

The mechanism of exchange tunneling, suggested by Appelbaum and
Anderson, was originally built into a  Hamiltonian with
constant, a priori unknown, exchange couplings to the spin. While this
approach relied on an exchange tunneling term connecting the two electrodes
(see the model \eqref{eq:hamilton} below), a parallel
development by S\'{o}lyom and Zawadowski\cite{Solyom68} suggested that
the conductance anomaly should arise from the energy dependent
renormalization of the local density of states, induced by Kondo spins
near one of the electrodes {\em not} directly coupled to the other side.
Initially the two proposals disagreed even in the sign of the change
in conductance, but these differences were resolved in a later work by
Appelbaum and Brinkman\cite{Appelbaum70}, using Zawadowskis
alternative approach to tunneling\cite{Zawadowski67,Appelbaum69}. The
final reconciliation of these ideas came with the work by
Ivezi\'{c}\cite{Ivezic75}, taking the nonequilibrium Keldysh approach
to tunneling, developed earlier by Caroli {\it et al.}\cite{Caroli71}.
The fact that the electrodes were out of mutual equilibrium was taken
into account, and it was demonstrated that the earlier (essentially
equilibrium) treatments\cite{Solyom68,Appelbaum70} were only correct
in cases when the impurity was located nearby one of the
electrodes, whereas an impurity somewhere in the middle of the barrier
constitutes a true nonequilibrium problem.  A good review of these
earlier works may be found in Ref.~\onlinecite{Wolf85}.

With the discovery of the Kondoesque tunneling anomaly in various
types of quantum dots, this venerable problem has recently been
revived. Meanwhile, experimenting with quantum dots offers much better
control over the parameters defining the problem, and this in turn
allows for a more systematic study of the physics underlying the
conductance anomaly. Since the zero-bias anomaly arises only in
Coulomb blockade valleys corresponding to an {\it odd} number of
electrons occupying the dot, the effective local moment can be
ascribed to a single electron in the uppermost energy level. In this
way the 'magnetic impurity' spreads over the entire dot and there is
no confusion as to whether the impurity is located close to, or even
residing in, one of the electrodes or whether one should average over
many impurities, issues which were all very important in the
conventional tunnel junctions mentioned above. 
In particular,  there is no reason to believe that
the dot-spin should be equilibrated with one particular lead,
which was pointed out by Ivezi\'{c}\cite{Ivezic75} to constitute a
true nonequilibrium problem.  Within the Anderson model, this problem
has been studied using an equations-of-motion technique combined with
the non-crossing approximation\cite{Meir93, Wingreen94, Hettler94,
Norlander99, Plihal00, Krawiec02}
and by means of nonequilibrium perturbation theory, expanding in the
hybridization strength\cite{Sivan96}. Other works have taken the Kondo
model as their starting point, but have mostly
focused on the effects of an applied ac-bias in the case of zero
magnetic field\cite{Goldin98,Kaminski99}.

A complete theory of the Kondo dot in a nonequilibrium stationary
state, i.e. in the presence of a finite current flowing through the
dot, does not exist yet. Most of the methods which have proven so successful
in dealing with the equilibrium Kondo problem, appear to have no
trivial extension to a nonequilibrium situation.  Schiller
{\it et al.}\cite{Schiller95} have successfully applied bosonization
techniques, originally devised by Emery and Kivelson\cite{Emery92} in
their solution of the two-channel Kondo problem, to calculate a number
of observables near a certain Toulouse point in the presence
of a finite bias voltage.  It is unclear,
  however, to what extent these results apply to the generic Kondo
  model.  Konik, Saleur and Ludwig\cite{Konik01} started from the
  Bethe Ansatz solution in equilibrium to construct {\em approximate}
  scattering states, in the presence of a finite voltage. To what
  extent their approximations and boundary conditions are valid is
  not obvious to us. To our knowledge, even the perturbation
theory in a true nonequilibrium situation has not yet been worked out
to leading logarithmic order in the presence of a magnetic field. This
latter task will be taken on in the present paper, which presents a
detailed analysis of the nonequilibrium perturbation theory to leading
logarithmic order, generalizing Appelbaum's result to the case where
the spin is not equilibrated with one of the leads.

Our aim here is to understand the physical processes governing the
nonequilibrium situation and to formulate a starting point for the
resummation of the leading logarithmic terms in perturbation theory.
Even in the perturbative regime, when magnetic fields, voltages or
temperatures are large compared to $T_K$, such a resummation is
necessary not only to recover the correct universal scaling behavior
but also to be able to fit experiments quantitatively\cite{Rosch03a}.
In some of the early works\cite{Appelbaum67b,Solyom68}, already the
perturbation theory was resummed using approximations of Nagaoka
\cite{Nagaoka65}, Abrikosov\cite{Abrikosov65} or Suhl and
Wong\cite{Suhl65}, mainly as an attempt to
capture the strong coupling regime.  As for the quantum dots, Kaminski
{\it et al.}\cite{Kaminski99} first suggested a poor mans scaling
method to deal with this problem at finite voltage, but this approach
did not encompass the case of a finite magnetic field and therefore
avoided the conceptual problem connected to the calculation of the
magnetization.  Recently, the present authors suggested a different
poor man's scaling approach\cite{Rosch03a},
which effectively resums the logarithmic corrections and recovers
one-parameter scaling, even in the presence of a magnetic field. As an
important check on this approach, it was crucial that an expansion in
bare parameters would correctly reproduce the leading logarithmic
corrections found from perturbation theory, and indeed this was found
to be the case. Main results of the nonequilibrium perturbation theory
presented here were briefly stated in Ref.~\onlinecite{Rosch03a}.
The details of our scaling approach will be provided in a subsequent
publication\cite{Rosch03b}.

The remaining parts of the paper are organized as follows. In section
\ref{model} we define our model and briefly review
the pseudo fermion formalism to be used throughout the paper. Section
\ref{diagrammatics} reviews the diagrammatic rules of nonequilibrium
perturbation theory using the Keldysh formalism, and introduces a few
basic building blocks from which all subsequent diagrams will be
constructed. In section \ref{nonequilmagnetization} we solve a quantum
Boltzmann equation to find the nonequilibrium distribution function
for the local spin, and demonstrate that this leads to rather dramatic
effects of the bias-voltage on the local magnetization. Finally, in
section \ref{tuncurrent} we derive an expression for the tunneling
current which takes the nonequilibrium magnetization as its input. 
Section \ref{discussion} contains a summary and discussion of our results.

\section{The Model}\label{model}

We consider the Hamiltonian \bea H&=&\sum_{\g,{\bf k},\s}(\e_{{\bf
    k}}-\mu_{\g})
c^{\dagger}_{\g{\bf k}\s}c_{\g{\bf k}\s}-g\mu_{B}B S_{z}\nonumber\\
& & + \!\sum_{\g,\g',{\bf k},{\bf k}',\s,\s'}\!\!\!\!\!J_{\g'\g}\,
\vec{S}\cdot\frac{1}{2}\,c^{\dagger}_{\g'{\bf k}'\s'}
\vec{\tau}_{\s'\s}c_{\g{\bf k}\s},\label{eq:hamilton} \eea where
$\mu_{L,R}=\pm eV/2$, $S_z$ is the spin-1/2 on the dot and $\vec{\tau}$ the
Pauli-matrices. We shall use the dimensionless coupling constants
$g_{i}=\Nf J_{i}$, where $\Nf$ is the density of states per spin
for the conduction electrons and $J_{LL}$, $J_{RR}$ and $J_{LR}=J_{RL}$
the real-valued exchange constants. For simple quantum dots in the Kondo
regime, which can be described by an Anderson model, the exchange
coupling constants are related by $g_{LR}^2=g_{LL} g_{RR}$. However,
in more complex situations, like in double dot systems in the Kondo
regime, no such relations  exist and we will therefore treat $g_{LR}$
as an independent parameter of our Hamiltonian (\ref{eq:hamilton}).
For notational convenience we employ the shorthand notation
$\gd=(\gll+\grr)/2$, and unless specifically stated otherwise, we will
henceforth use units where $\hbar=k_{B}=g\mu_{B}=e=1$.

In order to proceed with a perturbative calculation, which includes
also the effects of a magnetic field, it is convenient to apply a
fermionic representation of the local spin operators. We choose here
Abrikosovs \cite{Abrikosov65} {\it pseudo fermion} representation,
which, in terms of fermionic charge-neutral spin 1/2 operators $f$,
reads \be {\vec S}=\frac{1}{2}\sum_{\al\al'} f^{\dagger}_{\al}{\vec
  \tau}_{\al\al'}f_{\al'}.  \ee Since only the singly occupied fermion
states have any physical relevance, this representation must be
supplemented by a projection onto this physical part of the Hilbert
space, effectively excluding doubly occupied and empty states. To this
end, the pseudo fermion is endowed with a chemical potential
$\lambda$ which is kept finite throughout the calculation, using a
grand-canonical ensemble average. As demonstrated, for equilibrium and
nonequilibrium situations, in Refs.~\onlinecite{Zawadowski69}
and~\onlinecite{Wingreen94}, respectively, the physically relevant, i.e.
{\it canonical} ensemble averaged, expectation value of an observable
${\cal O}$ is obtained as the limiting value \be \langle{\cal
  O}\rangle_{Q=1}=\lim_{\lambda\to\infty} \frac{\langle{\cal
    O}Q\rangle_{\lambda}} {\langle
  Q\rangle_{\lambda}},\label{eq:canonav} \ee where
$Q=\sum_{\al}f^{\dagger}_{\al}f_{\al}$ is the pseudo fermion number
operator. Note that in the common case, where the observable ${\cal
  O}$ has zero expectation value in the $Q=0$ ensemble, one may leave
out the $Q$-operator from the numerator in (\ref{eq:canonav}).

Since $\lambda$ enters as a chemical potential, thermal averages taken with a
finite $\lambda$ contain various powers of $e^{-\lambda/T}$. In particular
${\langle Q\rangle_{\lambda}}\sim e^{-\lambda/T}$ and therefore the limit of
$\lambda\to\infty$ in Eq. (\ref{eq:canonav}) effectively picks out the terms in
$\langle{\cal O}Q\rangle_{\lambda}$ which are also proportional to
$e^{-\lambda/T}$. Likewise, in calculating any thermal average of interest at
finite $\lambda$, one is allowed to retain only the terms of lowest order in
$e^{-\lambda/T}$.

\section{Keldysh Diagrammatics}\label{diagrammatics}
 
In setting up the nonequilibrium perturbation theory, we shall comply with the
conventions in Ref.~\onlinecite{Rammer86}. All Keldysh-space matrix propagators
are represented in the usual upper-right triangular form
\bea
\undl{G}=\left(\begin{array}{cc}
G^{R} & G^{K} \\
0     & G^{A} \\
\end{array}\right)\label{eq:matrix}
\eea
and the individual entries will be denoted by Latin indices. From this basic
Green function, one may obtain the usual {\it spectral}, {\it lesser} and
{\it greater} functions as
\bea
A&=&i(G^{R}-G^{A}),\\
G^{<}&=&(G^{K}-G^{R}+G^{A})/2,\\
G^{>}&=&(G^{K}+G^{R}-G^{A})/2,\\
{\rm Re}[G]&=&(G^{R}+G^{A})/2,
\eea
implying the $R$ superscript in the real part. A corresponding notation will be
used for self-energies, except that the spectral function will be replaced by the
{\it broadening} $\G=i(\Sigma^{R}-\Sigma^{A})$. We shall henceforth denote
conduction electron ($ce$), and pseudo fermion ($pf$) correlation functions by
capital Latin, and calligraphic letters, respectively.

\begin{center}
\begin{figure}
\includegraphics[width=\linewidth]{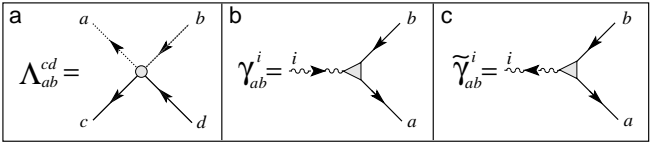}
\caption{\label{fig:vertices} (a) The bare 
conduction electron-pseudo fermion vertex. (b) The absorption,
and (c) the emission vertices. Propagators are dashed ($pf$) and full ($ce$)
lines. Note that the absorption, and emission vertices could equally well have
been drawn with $pf$-propagators.}
\end{figure}
\end{center}

Perturbation theory in terms of Keldysh matrix propagators involves the bare
four-point vertex and the {\it measurement} vertices depicted in
Fig.~\ref{fig:vertices}. These have the tensor-structure
\bea
\Lambda_{ab}^{cd}&=&\frac{1}{2}
\left(\delta_{ab}\tau^{1}_{cd}+\tau^{1}_{ab}\delta_{cd}\right),
\label{eq:vertex0}\\
\gamma^{1}_{ab}&=&{\tilde \gamma}^{2}_{ab}=\frac{1}{\sqrt{2}}\delta_{ab},
\label{eq:vertex1abs}\\
\gamma^{2}_{ab}&=&{\tilde \gamma}^{1}_{ab}=\frac{1}{\sqrt{2}}\tau^{1}_{ab}.
\label{eq:vertex1emi}
\eea
Diagrams should be interpreted with the usual zero-temperature Feynman rules,
including a prefactor $(iJ)^{n}(-1)^{{\text F}_{pf}+{\text F}_{ce}}$ to any
diagram of order $J^{n}$, having F$_{pf,ce}$ closed pseudo fermion, or
conduction electron loops. Both {\it flavors} propagate with spin-indices
$\al,\s =\pm 1$, to which we apply the convention that
${\overline \al}\equiv -\al$. Furthermore, the conduction electron Green
functions carry a lead-index $\g\in\{L,R\}$. The Einstein summation convention is
implied for all indices except where otherwise stated.

Since we assume the total system to have reached a steady state, all
Green functions depend only on one frequency. The bare $pf$ propagator
has the spectral function
\be
{\cal A}_{\al}(\w)=2\pi\delta(\w+\al B/2).\label{eq:barepfA}\ee
and corresponding retarded and advanced
Green functions. For the Keldysh component, we make the usual ansatz
that
\be
{\cal G}_{\al}^{K}(\w)=i{\cal A}_{\al}(\w)(2 n_{\al\la}(\w)-1),
\ee
which defines the $pf$-distribution-function $n_{\al\la}(\w)$. At zero
bias-voltage, where the spin is equilibrated with the conduction electrons,
$n_{\al\la}(\w)$ reduces to the Fermi function
$f(\w+\lambda)=1/(e^{(\w+\lambda)/T}+1)$.  We follow the convention of
Ref.~\onlinecite{Kadanoff62} and place the chemical potentials in
the distribution functions rather than in the spectral functions,
which prevents the parameter $\lambda$ from pervading the formulas.
The unprojected mean occupation numbers are given by the integral
\be
n_{\al\la}=\int^{\infty}_{-\infty}\!\!\frac{d\w}{2\pi}\, {\cal
  A}_{\al}(\w)n_{\al\la}(\w),\label{eq:occnumdef}
\ee
that is $n_{\al\la}=n_{\al\la}(-\al B/2)$ for bare ${\cal A}_{\al}$.
After taking the limit $\lambda\rightarrow\infty$ for a given observable and thus
performing the projection onto the physical spin-states, one can use the
constraint $n_{\uparrow}+n_{\downarrow}=1$ to write the local magnetization
as $M=2n_{\uparrow}-1$. At times, we shall also use the shorthand
$M_{\al,\lambda}=2n_{\al,\lambda}-1$ prior to projection.

The bare conduction electron Green functions depend on momentum as well
as frequency, but since the interaction and the pseudo fermions are local in
space, all internal lines in a Feynman diagram involve only propagation in time,
and we are allowed to work with local, momentum integrated, $ce$-Green functions.
Assuming a constant density of states in a band of width $2 D$, centered at zero,
the momentum integrated spectral function reads
\be
A(\w)=2\pi \Nf\Theta(D-|\w|),\label{eq:cespectralfct0}
\ee
with $\Nf=1/(2 D)$. Assuming the electrons in separate leads to be in thermal
equilibrium, the lead-dependent Keldysh Green function takes the form
\be
G^{K}_{\g}(\w)=-i A(\w) \tanh\left(\frac{\w-\mu_{\g}}{2 T}\right),
\ee
in terms of the chemical potentials $\mu_{L}=-\mu_{R}=V/2$.
The spectral function (\ref{eq:cespectralfct0}), implies a small real part
${\rm Re}[G^{R}(\w)]=
\frac{\Nf}{2}\ln\left(\frac{\w+D}{\w-D}\right)\approx\Nf\w/D$
which we can safely neglect for most of our discussion.

\subsection{Second Order Vertex Functions}\label{vertfct}

\begin{center}
\begin{figure}
\includegraphics[width=0.8\linewidth]{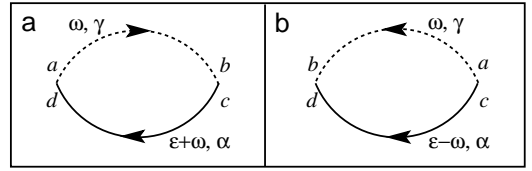}
\caption{\label{bubble} The conduction electron-pseudo fermion bubbles in
{\it Peierls} (a) and {\it Cooper} (b) channels. Conduction electron
propagators have lead-index $\g$, and pseudo fermion propagators carry
the spin index $\al$. Latin indices refer to the Keldysh-matrix
structure in (\ref{eq:matrix}).}
\end{figure}
\end{center}
The central objects in the diagrammatics of this problem are the $ce-pf$
bubbles depicted in Fig.~\ref{bubble}, insofar as they
carry the logarithmic integrals which eventually lead to the Kondo effect. The
bubbles in the {\it Peierls}- and {\it Cooper}-channels are evaluated
 as
\bea
\ind{\al\,}{\g}{{\undl \K}}{ba}{dc}(\e)
&=&\int\!\!\frac{d\w}{2\pi}\,
{\undl G}^{dc}_{\g}(\w+\e)\,
{\undl {\cal G}}^{ba}_{\al}(\w)\\
\ind{\al\,}{\g}{{\undl {\ovrl \K}}}{ba}{dc}(\e)
&=&\int\!\!\frac{d\w}{2\pi}\,
{\undl G}^{dc}_{\g}(\e-\w)\,
{\undl {\cal G}}^{ba}_{\al}(\w),
\eea
from which one can readily determine the various Keldysh-components by straightforward 
integrations:
\bea
{_{\al}}\K_{R}^{R}(\e)&=&{_{\al}}\K_{A}^{A}(\e)=0,\label{eq:Kfcts}\\
{_{\al}}\K_{A}^{R}(\e)
&=&i G^{R}(\e-\al B/2),\nonumber\\
{_{\al}}\K_{R}^{A}(\e)
&=&-i G^{A}(\e-\al B/2),\nonumber\\
{_{\al}}\!\!\!{^{\g}}\K_{K}^{R,A,K}(\e)
&=&i G^{R,A,K}_{\g}(\e-\al B/2)M_{\al,\lambda},\nonumber\\
{_{\al}}\!\!\!{^{\g}}\K_{R}^{K}(\e)
&=&-\frac{i}{2}G^{K}_{\g}(\e-\al B/2)\nonumber\\
& &-i\Nf\ln\left(
\frac{D^{2}}{(\e-\mu_{\g}-\al B/2)^{2}+T^{2}}\right),\nonumber\\
{_{\al}}\!\!\!{^{\g}}\K_{A}^{K}(\e)
&=&\frac{i}{2}G^{K}_{\g}(\e-\al B/2)\nonumber\\
& &-i\Nf\ln\left(
\frac{D^{2}}{(\e-\mu_{\g}-\al B/2)^{2}+T^{2}}\right),\nonumber
\eea
for the Peierls-channel, from which the bubbles in the Cooper-channel are
 obtained as
\bea
{_{\al}}{\ovrl \K}_{A}^{R}(\e)
&=&{_{\al}}{\ovrl \K}_{R}^{A}(\e)=0,\\
{_{\al}}{\ovrl \K}_{R,A}^{R,A}(\e)
&=&-{_{\al}}\K_{R,A}^{A,R}(-\e),\nonumber\\
{_{\al}}\!\!\!{^{\g}}{\ovrl \K}_{K}^{R,A,K}(\e)
&=&-{_{\al}}\!\!\!{^{{\ovrl \g}}}\K_{K}^{A,R,K}(-\e),\nonumber\\
{_{\al}}\!\!\!{^{\g}}{\ovrl \K}_{R,A}^{K}(\e)
&=&-{_{\al}}\!\!\!{^{{\ovrl \g}}}\K_{R,A}^{K}(-\e).\nonumber
\eea
In both channels the Kondo effect derives from the imaginary part of bubbles with
a $K$-component conduction electron together with a retarded, or an advanced
pseudo fermion. In such combinations, $\RE[{\cal G}_{\al}(\w)]\propto 1/\w$ is
convoluted with $\tanh(\e+\w/2T)$ to produce the logarithmic enhancement.

To work out higher order Feynman diagrams, it is convenient
to construct  the second-order renormalized vertex functions, obtained by
attaching a Keldysh vertex (\ref{eq:vertex0}) to each end of these
bubbles:
\bea
\ind{\al}{\g}{{\undl \I}}{ab}{cd}(\e)&=&\Lambda^{c'd}_{ab'}\,\,
\ind{\al\,}{\g}{{\undl \K}}{b'\!a'}{\!\!\!d'\!c'}(\e)
\Lambda^{cd'}_{a'b},\\
\ind{\al}{\g}{{\ovrl {\undl \I}}}{ab}{cd}(\e)
&=&\Lambda^{cd'}_{ab'}\,\,
\ind{\al\,}{\g}{{\undl {\ovrl \K}}}{b'\!a'}{\!\!\!d'\!c'}(\e)
\Lambda^{c'd}_{a'b}.\label{eq:vertex1}
\eea
Working out the contraction of Keldysh indices one may  organize all entries
in the Peierls channel as
\bea
{\undl \I}^{12}_{11}={\undl \I}^{22}_{12}=
{\undl \I}^{11}_{21}={\undl \I}^{21}_{22}&=&\frac{1}{4i}\I^{R},\nonumber\\
{\undl \I}^{11}_{12}={\undl \I}^{12}_{22}=
{\undl \I}^{22}_{21}={\undl \I}^{21}_{11}&=&\frac{1}{4i}\I^{A},\nonumber\\
{\undl \I}^{11}_{11}={\undl \I}^{22}_{22}=
{\undl \I}^{12}_{21}={\undl \I}^{21}_{12}&=&\frac{1}{4i}\I^{K},\nonumber\\
{\undl \I}^{11}_{22}={\undl \I}^{22}_{11}=
{\undl \I}^{12}_{12}={\undl \I}^{21}_{21}&=&0,\label{eq:Gamlist}
\eea
satisfying the symmetry ${\undl \I}^{cd}_{ab}=
{\undl \I}^{c{\ovrl d}}_{{\ovrl a}b} ={\undl \I}^{{\ovrl c}d}_{a{\ovrl b}}$,
and similarly in the Cooper-channel
\bea
{\ovrl {\undl \I}}^{12}_{22}={\ovrl {\undl \I}}^{22}_{12}=
{\ovrl {\undl \I}}^{11}_{21}={\ovrl {\undl \I}}^{21}_{11}&=&
\frac{1}{4i}{\ovrl \I^{R}},\nonumber\\
{\ovrl {\undl \I}}^{11}_{12}={\ovrl {\undl \I}}^{12}_{11}=
{\ovrl {\undl \I}}^{22}_{21}={\ovrl {\undl \I}}^{21}_{22}&=&
\frac{1}{4i}{\ovrl \I^{A}},\nonumber\\
{\ovrl {\undl \I}}^{11}_{22}={\ovrl {\undl \I}}^{22}_{11}=
{\ovrl {\undl \I}}^{12}_{21}={\ovrl {\undl \I}}^{21}_{12}&=&
\frac{1}{4i}{\ovrl \I^{K}},\nonumber\\
{\ovrl {\undl \I}}^{11}_{11}={\ovrl {\undl \I}}^{22}_{22}=
{\ovrl {\undl \I}}^{12}_{12}={\ovrl {\undl \I}}^{21}_{21}&=&0,
\label{eq:Gamlistcoop}
\eea
where ${\ovrl {\undl \I}}^{cd}_{ab}={\ovrl {\undl \I}}^{{\ovrl c}d}_{{\ovrl a}b}
={\ovrl {\undl \I}}^{c{\ovrl d}}_{a{\ovrl b}}$. For clarity, we have temporarily
suppressed the variables $\al$, $\g$ and $\e$, and written only the Keldysh
indices. Furthermore we have introduced new functions, $\I^{R,A,K}$ and
${\ovrl \I}^{R,A,K}$, with
\bea
\I^{R,A}&=&i\left(\K^{R,A}_{K}+\K^{K}_{A,R}\right),\nonumber\\
{\ovrl \I}^{R,A}&=&i\left({\ovrl \K}^{R,A}_{K}
                       +{\ovrl \K}^{K}_{R,A}\right),\nonumber\\
\I^{K}&=&i\left(\K^{K}_{K}+\K^{R}_{A}+\K^{A}_{R}\right),\nonumber\\
{\ovrl \I}^{K}&=&i\left({\ovrl \K}^{K}_{K}
                       +{\ovrl \K}^{R}_{R}
                       +{\ovrl \K}^{A}_{A}\right),
\eea
satisfying the relation:
$_{\al}^{\g}{\ovrl \I}^{R,A,K}(\e)=
- _{\al}^{{\ovrl \g}}\I^{A,R,K}(-\e).$

From their definitions, these new functions $\I^{R,A,K}$ are seen to have poles
in the lower ($R$), the upper ($A$), or both ($K$) half-planes, and to
satisfy that $(\I^{R}(\e))^{\ast}=\I^{A}(\e)$. To be specific, one finds from
(\ref{eq:Kfcts}):
\bwt
\bea
_{\al}^{\g}\I^{R}(\e)&=&
-M_{\al,\lambda}G^{R}_{\g}(\e-\al B/2)-
\frac{1}{2}G^{K}_{\g}(\e-\al B/2)+
\Nf\ln\left(\frac{D^{2}}{(\e-\mu_{\g}-\al B/2)^{2}+
T^{2}}\right),\\
_{\al}^{\g}\I^{K}(\e)&=&
iA(\e-\al B/2)\left\{1+
M_{\al,\lambda}\tanh\left(\frac{\e-\mu_{\g}-\al B/2}{2 T}\right)\right\}.
\eea
The real, and the imaginary part of $\I^{R}$ are
\bea
\RE\left[_{\al}^{\g}\I(\e)\right]&=&
\Nf\ln\left(\frac{D^{2}}{(\e-\mu_{\g}-\al B/2)^{2}+
T^{2}}\right)-
M_{\al,\lambda}\RE\left[G_{\g}(\e-\al B/2)\right],\label{eq:REvert}\\
\IM\left[_{\al}^{\g}\I(\e)\right]&=&
\frac{1}{2}A(\e-\al B/2)\left(M_{\al,\lambda}+\tanh
\left(\frac{\e-\mu_{\g}-\al B/2}{2T}\right)\right),\label{eq:IMvert}
\eea
while the {\it lesser} and {\it greater} components take the following form: 
\bea
_{\al}^{\g}\I^{<}(\e)&=&2iA(\e-\al B/2)(1-n_{\al,\lambda})
                        f(\e-\mu_{\g}-\al B/2),\label{eq:vertl}\\
_{\al}^{\g}\I^{>}(\e)&=&2iA(\e-\al B/2)n_{\al,\lambda}
                        (1-f(\e-\mu_{\g}-\al B/2)).\label{eq:vertg}
\eea
\ewt
Note that since $\RE[G(\w)]\approx \Nf \w/D$ for $\w\ll D$, contributions from
the last term in (\ref{eq:REvert}), of order $max[B,V]/D \ll 1$, can safely be
neglected.

The real and imaginary parts (\ref{eq:REvert}, \ref{eq:IMvert}) satisfy the
Kramers-Kronig relation, and altogether their analytical properties allow us to
interpret $\I$ as genuine Keldysh Green functions describing the (time-) parallel
or anti-parallel propagation of conduction electrons and pseudo fermions. The
different components ($R,A,<,>$) of this mixed bubble could have been
written down immediately using the Langreth rules\cite{Langreth76} for analytical
continuation, and in fact this method is also very convenient for determining the
different components of the second order self-energy. For the third order
self-energy, however, we find the Keldysh matrix structure to be more convenient
when dealing with the large number of contractions appearing to this order. The
catalogues (\ref{eq:Gamlist}) and (\ref{eq:Gamlistcoop}) have been worked out for
this purpose and are used repeatedly in working out the different Keldysh
contractions appearing throughout the paper.


\section{Nonequilibrium Magnetization}\label{nonequilmagnetization}

\subsection{Pseudo Fermion Self-Energy}\label{pfselfen}

With these basic diagrammatic objects at hand, we can now readily evaluate the
Feynman-diagrams for the $pf$ self-energies shown in Fig. \ref{selfen}. It is,
however, essential to realize that the occupation functions $n_{\gamma \lambda}$
on the dot are completely undetermined to zeroth order of perturbation theory,
i.e. in the absence of a coupling to the leads\cite{Parcollet02}. These are
therefore kept as free parameters to be determined later. The first order $pf$
self-energy vanishes, since we have not included a Zeeman term for the conduction
electrons. Nevertheless, the corresponding Hartree term would be entirely real
and contribute only by a constant shift of the $pf$ energy levels (cf. appendix).

\subsubsection{Second order self-energy}

\begin{center}
\begin{figure}[b]
\includegraphics[width=\linewidth]{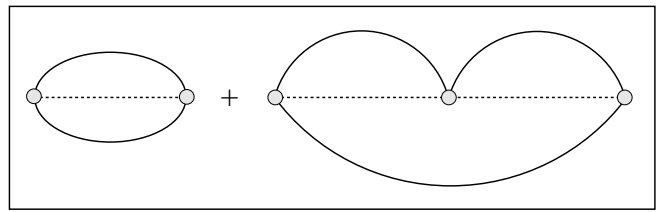}
\caption{\label{selfen} Unlabeled Feynman diagrams for the $pf$ self-energies.}
\end{figure}
\end{center}

Neglecting the Zeeman term for the conduction electrons, the $ab$-component
of the second-order self-energy from Fig.~\ref{selfen} is given as
\bea
{\undl \Sigma}^{ab(2)}_{\al\lambda}(\w)=
\frac{\theta_{\al\al'}}{8}J^{2}_{\g\g'}
\!\int\!\!\frac{d\e}{2\pi}
\,\ind{\al'}{\g}{\undl {\I}}{ab}{cd}(\e)
\,{\undl G}^{dc}_{\g'}(\e+\w),
\eea
where a summation over $ce$-spin has led to the tensor
\be
\theta_{\al\al'}=\frac{1}{2}\sum_{\s,\s'}
\tau^{i}_{\s\s'}\tau^{i}_{\al\al'}
\tau^{j}_{\s'\s}\tau^{j}_{\al'\al}=\delta_{\al\al'}+2\tau^{1}_{\al'\al}.
\label{eq:theta2}\ee
Contracting the Keldysh indices $c$ and $d$, using (\ref{eq:Gamlist}), and using
the analytical properties of $\I$, the $K$ component of this self-energy may be
written as
\bea
\Sigma^{K(2)}_{\al\lambda}(\w)&=&
\frac{\theta_{\al\al'}}{16i}J^{2}_{\g\g'}
\!\int\!\!\frac{d\e}{2\pi}\left[
{_{\al'}}\!\!\!\!{^{\g}}\I^{<}(\e)
G_{\g'}^{>}(\e+\w)\right.\\
&&\left.\hspace*{20mm}+
{_{\al'}}\!\!\!\!{^{\g}}\I^{>}(\e)
G_{\g'}^{<}(\e+\w)\right].\nonumber
\eea
Similarly, from the retarded and advanced components of the self-energy, the
imaginary part is found to be
\bea
\G^{(2)}_{\al\lambda}(\w)&=&
\frac{\theta_{\al\al'}}{16}J^{2}_{\g\g'}
\!\int\!\!\frac{d\e}{2\pi}\left[
{_{\al'}}\!\!\!\!{^{\g}}\I^{<}(\e)
G_{\g'}^{>}(\e+\w)\right.\\
&&\left.\hspace*{20mm}-
{_{\al'}}\!\!\!\!{^{\g}}\I^{>}(\e)
G_{\g'}^{<}(\e+\w)\right],\nonumber
\eea
and the lesser component
\bea
\Sigma^{<(2)}_{\al\lambda}(\w)&=&
\frac{\theta_{\al\al'}}{16i}J^{2}_{\g\g'}
\!\int\!\!\frac{d\e}{2\pi}
{_{\al'}}\!\!\!\!{^{\g}}\I^{>}(\e)
G_{\g'}^{<}(\e+\w).\nonumber\\
\eea

Using the identities (\ref{eq:vertl}, \ref{eq:vertg}) it is now straightforward
to carry out the integral over $\e$, together with the summation over $\g$, $\g'$
and $\al'$, to find that
\be 
\G^{(2)}_{\uparrow}(-B/2)=
\frac{\pi}{4}\left[
N^{(2)}_{\uparrow\uparrow}+
N^{(2)}_{\uparrow\downarrow}+
R^{(2)}
\right]\label{eq:gamma2}
\ee
and
\be
\Sigma^{<(2)}_{\uparrow\lambda}(-B/2)=\frac{i\pi}{4}\left[
N^{(2)}_{\uparrow\uparrow}n_{\uparrow,\lambda}+
N^{(2)}_{\uparrow\downarrow}n_{\downarrow,\lambda}\right],
\label{eq:sigmaless2}
\ee
where the coefficients are given by
\bea
N^{(2)}_{\uparrow\uparrow}&=&
\glr^{2}V\coth(V/2T)+(\gll^{2}+\grr^{2})T,\label{eq:Nuu2}\\
N^{(2)}_{\uparrow\downarrow}&=&
2(\gll^{2}+\grr^{2})B(1+N(B))\label{eq:Nud2}\\
&&+2\glr^{2}\left[(B+V)\left(1+N(B+V)\right)\right.\nonumber\\
&&\hspace*{9mm}\left.+(B-V)\left(1+N(B-V)\right)\right],\nonumber\\
R^{(2)}&=&-2(\gll^{2}+\grr^{2}+2\glr^{2})B\label{eq:R2},
\eea
and $N(\w)=1/(e^{\w/T}-1)$ is the Bose-function. While the coefficients 
$N_{\al\al'}$ do not depend on $\lambda$, $R^{(2)}$ is in fact given by
$(n_{\downarrow,\lambda}-1)8g^{2}B$. Since, however, the factor
$n_{\downarrow,\lambda}$ vanishes in the limit of $\lambda\to\infty$ and leaves
behind a term which remains finite, we can simply omit this term and employ the
identity (\ref{eq:R2}) for $R$.

\subsubsection{Third order self-energy}

The third order self-energy diagram depicted in Fig.~\ref{selfen} gives rise to
two different terms, corresponding to two different orientations on the
$ce$-loop. One involves two Cooper-bubbles and the other, two
Peierls-bubbles, and the combined $ab$-component translates to
\bwt
\bea
{\undl \Sigma}^{ab(3)}_{\al\lambda}(\w)&=&
\frac{i}{64}J_{\g\g'}J_{\g'\g''}J_{\g''\g}
\!\int\!\!\frac{d\e}{2\pi}\,
\left[
\theta^{P}_{\al\al'\al''}
\Lambda_{ab'}^{c'\!d}
\ind{\al''\!}{\g\,}{{\undl \K}}{b'\!a''}{\!\!\!\!\!d''\!c'}
(\e)\,
\Lambda_{a''b''}^{c''d''}\,
\ind{\al'\!}{\g'}{{\undl \K}}{b''\!a'}{\!\!\!\!\!d'\!c''}
(\e)\,
\Lambda_{a'\!b}^{cd'}\,
\,{\undl G}^{dc}_{\g''}(\e+\w)
\right.\nonumber\\
&&\hspace*{33mm}+\left.
\theta^{C}_{\al\al'\al''}
\Lambda_{ab'}^{cd'}\,
\ind{\al''\!}{\g\,}{{\ovrl {\undl \K}}}{b'\!a''}{\!\!\!\!\!d'\!c''}
(\e)\,
\Lambda_{a''\!b''}^{c''\!d''}\,
\ind{\al'\!}{\g'}{{\ovrl {\undl \K}}}{b''\!a'}{\!\!\!\!\!d''\!c'}
(\e)\,
\Lambda_{a'\!b}^{c'\!d}\,
\,{\undl G}^{dc}_{\g''}(\e-\w)
\right],\label{eq:selfen3ab}
\eea
where a summation over $ce$-spin has produced the tensor
\bea
\theta^{P}_{\al\al'\al''}&=&
-2i\epsilon_{ijk}\tau^{i}_{\al\al'}\tau^{j}_{\al'\al''}\tau^{k}_{\al''\al}
=4[\delta_{\al\al'} \tau^{1}_{\al'\al''}
  +\delta_{\al\al''}\tau^{1}_{\al\al'}
  +\delta_{\al'\al''}\tau^{1}_{\al\al'}]\nonumber
\eea
and $\theta^{C}_{\al\al'\al''}=-\theta^{P}_{\al\al'\al''}$. A contraction of the
relevant Keldysh indices allows one to express (\ref{eq:selfen3ab}) as a sum of
products of two vertex functions. Using again the relations
derived in section~\ref{vertfct}, together with the analytical properties of
$\I$, the $K$ component of the third order self-energy may be written as
\bea
\Sigma^{K(3)}_{\al\lambda}(\w)&=&
\frac{1}{128i}\theta^{P}_{\al\al'\al''}J^{3}_{\g\g'\g''}
\!\int\!\!\frac{d\e}{2\pi}
\RE\left[{_{\al''\!}}\!\!\!\!{^{\g\,}}\I(\e)\right] \left[
{_{\al'\!}}\!\!\!\!{^{\g'\!}}\I^{<}(\e)
G_{\g''}^{>}(\e+\w)+
{_{\al'\!}}\!\!\!\!{^{\g'\!}}\I^{>}(\e)
G_{\g''}^{<}(\e+\w)\right]+(V \leftrightarrow -V),\nonumber
\eea
where we have introduced the shorthand
$J^{3}_{\g\g'\g''}=J_{\g\g'}J_{\g'\g''}J_{\g''\g}$. The imaginary part takes
the similar form
\bea
\G^{(3)}_{\al\lambda}(\w)&=&
\frac{1}{128}\theta^{P}_{\al\al'\al''}J^{3}_{\g\g'\g''}
\!\int\!\!\frac{d\e}{2\pi}
\RE\left[{_{\al''\!}}\!\!\!\!{^{\g\,}}\I(\e)\right]
\left[
{_{\al'\!}}\!\!\!\!{^{\g'\!}}\I^{<}(\e)
G_{\g''}^{>}(\e+\w)-
{_{\al'\!}}\!\!\!\!{^{\g'\!}}\I^{>}(\e)
G_{\g''}^{<}(\e+\w)\right]
+(V \leftrightarrow -V).\nonumber
\eea
and the lesser component reads
\bea
\Sigma^{<(3)}_{\al\lambda}(\w)=
\frac{1}{128i}\theta^{P}_{\al\al'\al''}J^{3}_{\g\g'\g''}
\!\int\!\!\frac{d\e}{2\pi}
\RE\left[{_{\al''\!}}\!\!\!\!{^{\g\,}}\I(\e)\right]
{_{\al'\!}}\!\!\!\!{^{\g'\!}}\I^{>}(\e)
G_{\g''}^{<}(\e+\w)
+(V \leftrightarrow -V).
\eea
Comparing to the second order result, the basic difference is the presence of the
extra real part of the vertex-function, which provides the logarithmic
enhancement underlying the Kondo effect. 

Again, the relevant integrations and summations are carried out
using the specific form of the various components of $\I$, and the
result is expressed just as in Eqs. (\ref{eq:gamma2}) and
(\ref{eq:sigmaless2}), but with new third order coefficients
\bea
N^{(3)}_{\uparrow\uparrow}&=&
2\glr^{2}\gd\left\{
\left[T+(V+B)\coth\left(\frac{V}{2T}\right)\right]\ln\frac{D}{|V+B|}+
\left[T+(V-B)\coth\left(\frac{V}{2T}\right)\right]\ln\frac{D}{|V-B|}
\right\}\nonumber\\
&&+2(\gll^{3}+\grr^{3})T\ln\frac{D}{|B|},\label{eq:Nuu3}\\
N^{(3)}_{\uparrow\downarrow}&=&
4\glr^{2}\gd\left\{\hspace*{2mm}
(1+N(B+V))\left[(B+V)\ln\frac{D}{|B+V|}
  +B\ln\frac{D}{|B|}+V\ln\frac{D}{|V|}\right]\right.\nonumber\\
&&\hspace*{13.5mm}+(1+N(B-V))\left[(B-V)\ln\frac{D}{|B+V|}
  +B\ln\frac{D}{|B|}-V\ln\frac{D}{|V|}\right]\label{eq:Nud3}\\
&&\left.\hspace*{13.5mm}+(1+N(B))\left[(B+V)\ln\frac{D}{|B+V|}
  +(B-V)\ln\frac{D}{|B-V|}\right]\,
\right\}+4(\gll^{3}+\grr^{3})B(1+N(B))\ln\frac{D}{|B|},\nonumber
\\
R^{(3)}&=&
-8 \glr^{2}\gd\left\{
 (B+V)\ln\frac{D}{|B+V|}
+(B-V)\ln\frac{D}{|B-V|}
+B\ln\frac{D}{|B|}\right\}
-4(\gll^{3}+\grr^{3})B(1+N(B))\ln\frac{D}{|B|},\nonumber \\
\label{eq:R3}
\eea
\ewt
using the shorthand $\ln(D/|x|)=\ln(D/\sqrt{x^{2}+T^{2}})$. As was the case for
the second order terms, only $R^{(3)}$ depends on $\lambda$, but in such a way
that one can safely take the limit $\lambda\to\infty$ which leads to
(\ref{eq:R3}).

Logarithmic correction of the form $\ln(D/|B|)$, $\ln(D/|V|)$ or
$\ln(D/|V\pm B|)$, as in Eqs.~(\ref{eq:Nuu3}--\ref{eq:R3}), appear
throughout this paper. They are signatures of resonant scattering from one Fermi
surface to another. More precisely, a term like $\ln(D/|V|)$ shows that
resonant scattering from the left, to the right Fermi surface is prohibited due
to the difference $V$ in their electro-chemical potentials. Similarly, a
spin-flip process within the left lead is cut off by the required energy $B$,
leading to $\ln(D/|B|)$. Most interesting are probably the logarithms of the form
$\ln(D/|V-B|)$, characteristic for resonant scattering from the left, to the
right lead, where the energy mismatch of the two Fermi surfaces is compensated by
a spin-flip. These terms are especially important, insofar as they lead to
pronounced cusps at $V\sim B$ in physical quantities (see below).

\subsection{Solution of the Quantum Boltzmann Equation}\label{magn}

Having determined the $pf$ self-energy in terms of the unknown
{\it nonequilibrium} $pf$ occupation numbers, the Keldysh-component of the Dyson
equation provides a closed equation for $n_{\al\lambda}$. Since we assume the
system to be in a steady state, this equation is greatly simplified and may be
expressed as\cite{Rammer86}
\be
\G_{\al\lambda}(\w){\cal G}_{\al\lambda}^{K}(\w)=
{\cal A}_{\al}(\w)\Sigma^{K}_{\al\lambda}(\w),
\ee
or equivalently
\be
\G_{\al\lambda}(\w){\cal G}_{\al\lambda}^{<}(\w)=
{\cal A}_{\al}(\w)\Sigma^{<}_{\al\lambda}(\w),
\label{eq:qbeq}
\ee
which merely states that the collision integral in the quantum Boltzmann equation
has to vanish in a steady state situation.

The $pf$ spectral function appearing in this equation may be determined by
solving the retarded and advanced components of the Dyson equation and takes the
usual form
\be
{\cal A}_{\al}(\w)=
\frac{\G_{\al}(\w)}
{(\w-\RE[\Sigma_{\al}(\w)]+\al B/2)^{2}+(\G_{\al}(\w)/2)^{2}},\label{eq:renpfA}
\ee
where the shift $\RE[\Sigma]$ and the broadening $\G$ of the $pf$-energy levels
are determined from perturbation theory, including leading logarithmic
corrections. However, assuming that ${\cal A}_{\al}(\w)$ is non-zero for all
frequencies it can be divided out of Eq. (\ref{eq:qbeq}) which then takes the
simple form
\be
(\Sigma^{<}_{\al\lambda}(\w)-\Sigma^{>}_{\al\lambda}(\w))n_{\al\lambda}(\w)=
\Sigma^{<}_{\al\lambda}(\w),
\ee
or equivalently
\be
n_{\al\lambda}(\w)=
\left(1-\Sigma^{>}_{\al\lambda}(\w)/\Sigma^{<}_{\al\lambda}(\w)\right)^{-1}.
\label{eq:dfctid}
\ee
In this equation the self-energies $\Sigma^{<,>}_{\al\lambda}$ are
determined in perturbation theory as integrals involving the unknown
distribution function $n_{\al\lambda}(\w)$. However, in bare perturbation
theory, where there is no dressing of internal lines,
$n_{\al\lambda}(\w)$ is always multiplied by the {\it unrenormalized}
$pf$ spectral function (\ref{eq:barepfA}) inside the integrals, and
only the occupation {\it numbers} $n_{\al\lambda}\equiv
n_{\al\lambda}(-\al B/2)$ will appear in $\Sigma^{<,>}$. In our case, the quantum
Boltzmann equation can thus be solved without any
feedback from the retarded and advanced Dyson equations, and we can
focus our attention on the on-shell occupation {\it numbers}
$n_{\gamma \lambda}$.
\begin{center}
\begin{figure}[t]
\includegraphics[width=0.9\linewidth]{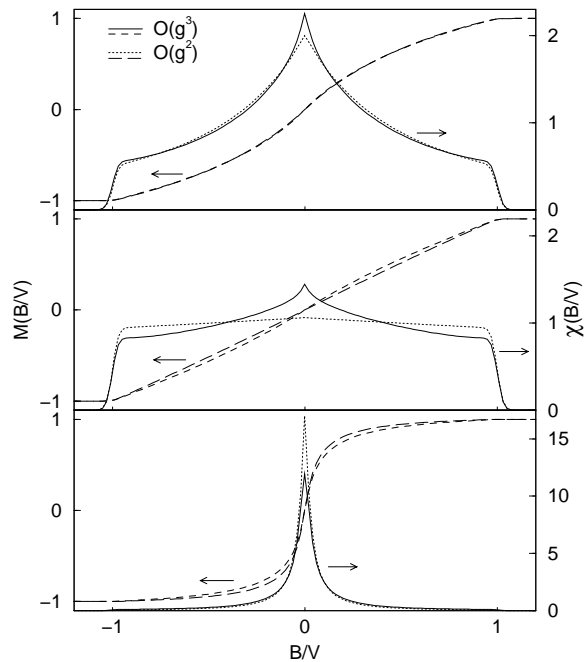}
\caption{\label{magnplotsB} Nonequilibrium magnetization (dashed and long-dashed
lines) and susceptibility (full and dotted lines) as functions of $B/V$, for
$T=10^{-2} V$ and $D=10^3 V$. Solid and dashed (dotted and long-dashed) lines
correspond to 3.(2.) order perturbation theory.
Upper panel: $\gll=\grr=\glr=0.05$.
Middle panel:  $\gll=\grr=0.05$ and $\glr=0.2$.
Lower panel:  $\gll=\grr=0.2$ and $\glr=0.05$
(note the different scale for $\chi$). Corrections from third order are more
pronounced for $g_{LR}^2 \neq g_{LL} g_{RR}$.}
\end{figure}
\end{center}

Dividing out the spectral function and setting $\w=-\al B/2$, Eq.
(\ref{eq:qbeq}) takes the form \be i\G_{\al\lambda}(-\al
B/2)n_{\al\lambda}=\Sigma^{<}_{\al\lambda}(-\al B/2).  \ee Since
$n_{\al\lambda}$, and thereby $\Sigma^{<}_{\al\lambda}$ vanishes in
the limit of $\lambda\to\infty$, the whole equation may be divided by
$\langle Q\rangle_{\lambda}$ and after this limit has been taken
  one can establish the same equation for the physical, projected,
occupation numbers.  Expressed in terms of the coefficients defined in
Eqs. (\ref{eq:gamma2}) and (\ref{eq:sigmaless2}), we arrive at the equation
\be
(N_{\up\up}+N_{\up\down}+R)n_{\up}=N_{\up\up}n_{\up}+N_{\up\down}n_{\down},
\label{eq:rateeq}
\ee 
which can be viewed as a rate-equation with the transition rates
$W_{\down\up}=\frac{\pi}{4}(N_{\up\down}+R)$ and
$W_{\up\down}=\frac{\pi}{4}N_{\up\down}$. This equation is readily
solved together with the constraint equation $n_{\up}+n_{\down}=1$ and
one finds that \be\label{nO}
n_{\up}=\frac{N_{\up\down}}{2N_{\up\down}+R}, \ee or expressed in
terms of the magnetization $M=n_{\up}-n_{\down}$ \be \label{mO}
M=\frac{-R}{2N_{\up\down}+R}.  \ee Note that to obtain the observable
magnetization to order $g^2 \ln(D)$, it is not sufficient to consider
only the on-shell occupations appearing in Eqs.~(\ref{eq:rateeq}--\ref{mO}). In
addition, one has to consider also contributions from the $pf$ spectral function
[see Eq.~(\ref{eq:occnumdef})], which are discussed in detail in the appendix.

Inserting the second order expressions (\ref{eq:Nuu2}--\ref{eq:R2}), one obtains
\bwt
\be
M(B,V)=
\frac{(\gll^{2}+\grr^{2}+2\glr^{2})B}
{(\gll^{2}+\grr^{2})B\coth\left(\frac{B}{2T}\right)
 +\glr^{2}\left[(B+V)\coth\left(\frac{B+V}{2T}\right)
                   +(B-V)\coth\left(\frac{B-V}{2T}\right)\right]},\label{mSec}
\ee
which, up to a factor of 2 in the definition of M, is exactly what was found in
Eq. (4) of Ref.~\onlinecite{Parcollet02}, where a rate-equation like
(\ref{eq:rateeq}) was solved using second order transition rates.
Including the third order corrections (\ref{eq:Nuu3}--\ref{eq:R3}), we obtain
\be
M={\cal T}/{\cal N}\label{magAll}
\ee
with
\bea {\cal T}&=&
4\glr^{2}\gd\left[(V+B)\ln\frac{D}{|V+B|}-(V-B)\ln\frac{D}{|V-B|}\right]
+B\sum_{\g}(g_{L\g}^{2}+g_{R\g}^{2})\left(1+2g_{\g\g}\ln\frac{D}{|B|}\right),
\label{eq:MpertT}\\
{\cal N}&=&\coth\left(\frac{B}{2T}\right)
\left\{
2\glr^{2}\gd\left[(V+B)\ln\frac{D}{|V+B|}-(V-B)\ln\frac{D}{|V-B|}\right]
+B\sum_{\g}g_{\g\g}^{2}\left(1+2g_{\g\g}\ln\frac{D}{|B|}\right)\right\}
\nonumber\\
&&+\coth\left(\frac{V+B}{2T}\right)\glr^{2}
\left[(V+B)\left(1+2\gd\ln\frac{D}{|V+B|}\right)
  +2V\gd\ln\frac{D}{|V|}
  +2B\gd\ln\frac{D}{|B|}\right]\nonumber\\
&&+\coth\left(\frac{V-B}{2T}\right)\glr^{2}
\left[(V-B)\left(1+2\gd\ln\frac{D}{|V-B|}\right)
  +2V\gd\ln\frac{D}{|V|}
  -2B\gd\ln\frac{D}{|B|}\right].\label{eq:MpertN} \eea \ewt 
As expected, the Kondo effect reveals itself in logarithmic enhancements,
and interestingly enough the logarithmic corrections to $M$ come as
$[1+g\ln(..)]$ rather than $[1+g^{2}\ln(..)]$, which is found in
equilibrium. In the limit where $T\gg V$, the logarithmic corrections
from ${\cal T}$ and ${\cal N}$ cancel and we recover the usual thermal
magnetization $M=\tanh(B/2T)$. In the highly asymmetric case, where
$\gll\gg\grr,\glr$, a similar cancellation takes place and one finds
again that $M=\tanh(B/2T)$.

In the appendix, we demonstrate how the observable magnetization
receives additional corrections of order $g^2\ln(..)$, arising from
self-energy corrections to the $pf$ spectral function. Nevertheless, in the
case where $V\gg T$, such additional corrections are subleading.

\begin{center}
\begin{figure}[t]
\includegraphics[width=0.95\linewidth]{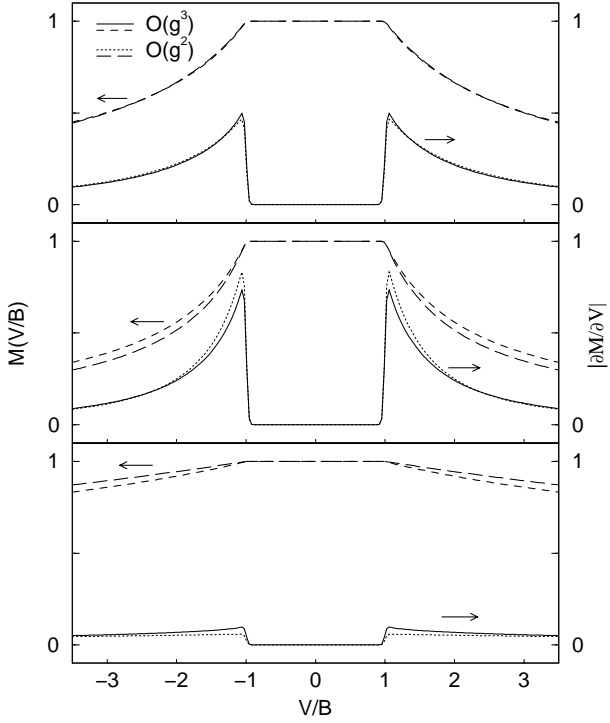}
\caption{\label{magnplotsV} Nonequilibrium magnetization as a function of $V/B$,
for $T=10^{-2} B$ and $D=10^3 B$. The two lower curves in each panel show the
absolute value of the slope, i.e. $\left|\frac{\partial M}{\partial V}\right|$.
This quantity enters the conductance shown in Fig.~\ref{condplotsV}. Solid
(dotted) lines correspond to 3.(2.) order perturbation theory.
Upper panel:  $\gll=\grr=\glr=0.05$.
Middle panel:  $\gll=\grr=0.05$ and $\glr=0.2$.
Lower panel:  $\gll=\grr=0.2$ and $\glr=0.05$.}
\end{figure}
\end{center}

In Fig.~\ref{magnplotsB} we plot the magnetization and the
corresponding susceptibility $\frac{\partial M}{\partial B}$ as
functions of $B/V$ for $T\ll V$, while Fig.~\ref{magnplotsV}
investigates the $V$ dependence of $M$ and $\partial M/\partial V$,
which will both influence the conductance. Roughly speaking, the
magnetization curve resembles the usual thermal magnetization with $V$
replacing $T$, and the impurity-spin becomes polarized only when $B$
exceeds $V$.  However, structures close to $V\sim B$ are much sharper
and obtain quite different logarithmic corrections compared to the
equilibrium case.  The exchange-correlations change the slopes of the
magnetization, which is clearly seen in the susceptibility
(Fig.~\ref{magnplotsB}) and in $\partial M/\partial V$ (Fig.~\ref{magnplotsV})
which become sharper spiked when including third order perturbations. In the
limit of $V\gg\max(B,T)$, one may expand to find
\bea \label{magVlarge}
\lefteqn{\hspace*{-16mm}M\approx\frac{B}{V}\left[
    \left(\frac{g_{L\g}^{2}+g_{R\g}^{2}}{2\glr^{2}}\right)
    \left(1+2g_{\g\g}\ln\frac{D}{|B|}\right)\right.}\nonumber\\
&&\left.\hspace*{6mm}
  -\left(\frac{\gll^{2}+\grr^{2}}{2\glr^{2}}\right)4\gd\ln\frac{D}{|V|}\right],
\label{eq:approxmagn}
\eea
which simplifies to $M=(2B/V)(1+2g\ln|V/B|)$ when $\gll=\grr=\glr$.
Note that we still imply the cut-off by $T$ meaning that $|B|$ and
$|V|$ in the argument of the logarithm should be interpreted as
$\sqrt{B^2+T^2}$ or $\sqrt{V^2+T^2}$, respectively. 

It is interesting to note that the structure of logarithmic
corrections is rather special in the case when the Hamiltonian
(\ref{eq:hamilton}) is derived from an underlying Anderson model by
means of a Schrieffer-Wolf transformation. In this case the exchange
couplings are related as $\glr^{2}=\gll\grr$ (cf. e.g.
Ref.~\onlinecite{Kaminski99}) and for such realizations of
the model, (\ref{eq:approxmagn}) simplifies to
\be
M\approx\frac{B}{V}\left\{1+\frac{\gll^{2}+\grr^{2}}{2\gll \grr}
\left[1+4\gd\ln\frac{|V|}{|B|}\right]\right\},\label{eq:AMmagn}
\ee
disclosing a partial
cancellation of logarithms, in the sense that the bandwidth, $D$, no
longer enters the logarithmic correction [the same is also valid  for the 
nonlinear magnetization (\ref{magAll})]. This explains why the plots of the
magnetization in the upper panels in Figs. \ref{magnplotsB} and
\ref{magnplotsV} hardly differ from second to third order, while more pronounced
effects are seen in the lower panels where $g_{LR}^2\neq g_{LL} g_{RR}$.

From Eq.~(\ref{eq:AMmagn}) we see that the peak at $B=0$ in the
magnetic susceptibility, seen in Fig.~\ref{magnplotsB}, grows as $\ln(|V|/T)$ for
$T\ll V$. Also, from (\ref{eq:approxmagn}) we learn that
the high voltage tails in Fig.~\ref{magnplotsV} fall off as $1/V$. The derivative
$\frac{\partial M}{\partial V}$, shown in Fig.~\ref{magnplotsV}, displays a
peak near $V=B$ already in second order perturbation theory which is then
slightly enhanced or reduced by the third order correction, depending on the
relative size of $\gd$ and $\glr$. As we shall see in section \ref{tuncurrent},
these features turn out to have a marked influence on the conductance.

For arbitrary $T$ and $V$, the susceptibility at $B=0$ takes the following form: 
\bwt
\bea
\chi&=&
\frac{
2\glr^{2}\left(1+2\gd\ln\frac{D}{T}+2\gd\ln\frac{D}{|V|}\right)
+\sum_{\g}g_{\g\g}\left(g_{\g\g}+2g_{\g\g}^2\ln\frac{D}{T}+
                                   2\glr^{2}\ln\frac{D}{|V|}\right)
}
{
2V\coth\left(\frac{V}{2T}\right)\glr^{2}\left(1+4\gd\ln\frac{D}{|V|}\right)
+2 T\sum_{\g}g_{\g\g}\left(g_{\g\g}+2g_{\g\g}^2\ln\frac{D}{T}+
                                      2\glr^{2}\ln\frac{D}{|V|}\right)
}.
\label{eq:chi}
\eea
\ewt
In the limit of $T\gg V$ the logarithms all take the form $\ln(D/T)$; the
corrections in numerator and denominator cancel, and we are left with the usual
Curie law $\chi=1/2T$. In the extreme nonequilibrium situation, however, where
$V\gg T$, the corrections no longer cancel and we are left with a complicated
fraction times $1/V$.

It is important to note that numerator and denominator in
(\ref{eq:chi}) have a rather different structure of logarithmic
corrections, e.g. the $g_{LR}^2$ term in the numerator receives corrections
of the form $2 g_{d}[\ln(D/T) +\ln(D/|V|)]$ whereas the corresponding term in
the denominator has the form $4 g_{d}\ln(D/|V|)$.  This observation was
the basis of our claim in Ref.~\onlinecite{Rosch03a} that the
perturbative renormalization group has to be formulated in terms of
coupling functions which depend on the energy of the incoming
electron. This will be explained in more detail in a forthcoming 
publication\cite{Rosch03b}.

\section{Nonlinear Tunneling Current}
\label{tuncurrent}

The current operator measuring the charge flow from left to right lead is found
from the equation of motion for the charge density in the left lead:
\be
\partial_{t}n_{L}=i[H, n_{L}].
\ee
The expectation value at time $t$ is
\bea
j_{L}=i J_{LR}\sum_{\stackrel{{\bf k},{\bf k}'}{\s,\s'}}
\left\langle\,\vec{S}(t)\cdot
[c^{\dagger}_{L{\bf k}'\s'}(t)\vec{\tau}_{\s'\s}c_{R{\bf k}\s}(t)
-L\leftrightarrow\!R\,]
\right\rangle,\nonumber
\eea
and by defining a two-particle Keldysh contour-ordered correlation function
\bea
\lefteqn{D_{LR}(\tau,\tau')=}\\
&&(-i)^{2}\sum_{{\bf k},{\bf k}'}
\left\langle T_{c_{K}}\left\{
c^{\dagger}_{L{\bf k}'\s'}(\tau)\frac{\vec{\tau}_{\s'\s}}{2}c_{R{\bf k}\s}(\tau)
\cdot\vec{S}(\tau')
\right\}\right\rangle\nonumber,
\eea
the current may be expressed simply by the imaginary part of the Keldysh
component
\be
j_{L}=J_{LR} \IM[D^{K}_{LR}(t,t)].\label{eq:currentdef}
\ee
\begin{center}
\begin{figure}[t]
\includegraphics[width=\linewidth]{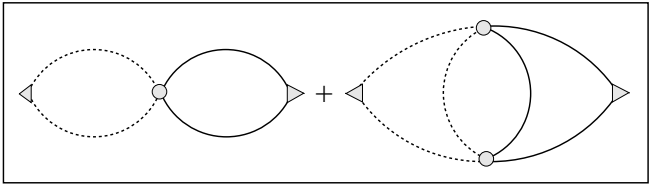}
\caption{\label{current} Unlabeled Feynman diagrams for $D_{LR}$. Triangles
denote the bare measurement vertices.}
\end{figure}
\end{center}

The lowest order contribution to this correlation function is
obtained from the first Feynman diagram in Fig.~\ref{current}, which translates
to
\bea
\lefteqn{D^{K(1)}_{LR}(t,t)=-\frac{\theta_{\al\al'}}{8 i}J_{LR}
\!\int\!\!\frac{d\e}{2\pi}
\!\int\!\!\frac{d\w}{2\pi}
\!\int\!\!\frac{d\W}{2\pi}}\\
&&\!\!\times
{\tilde \gamma}^{1}_{a'b'}
{\cal G}^{ba'}_{\al'}(\w)
{\cal G}^{b'a}_{\al}(\w+\W)
\Lambda_{ab}^{cd}
G^{d'c}_{R}(\e)
G^{dc'}_{L}(\e+\W)
\gamma^{2}_{c'd'}\nonumber
\eea
with measuring vertices ${\tilde \gamma}^{1}$ and $\gamma^{2}$ defined in
(\ref{eq:vertex1abs}, \ref{eq:vertex1emi}) and the spin-tensor from
(\ref{eq:theta2}). Contracting the Keldysh indices and expressing things in
terms of the vertex function $\I$, this simplifies to
\bea
\lefteqn{D^{K(1)}_{LR}(t,t)=\frac{\theta_{\al\al'}}{128 i}J_{LR}
\!\int\!\!\frac{d\e}{2\pi}}\\
&&\!\!\times\left[
 {_{\al}}\!\!\!\!{^{L}}\I^{K}(\e)
 {_{\al'\!}}\!\!\!\!{^{R}}\I^{R}(\e)
+{_{\al}}\!\!\!\!{^{L}}\I^{A}(\e)
 {_{\al'\!}}\!\!\!\!{^{R}}\I^{K}(\e)
\right],\nonumber
\eea
which in turn leads to
\bea
j_{L}^{(2)}&=&\frac{\theta_{\al\al'}}{32i}J_{LR}^{2}
\!\int\!\!\frac{d\e}{2\pi}\,
\IM\left[{_{\al}}\!\!\!\!{^{R}}\I(\e)\right]
{_{\al'\!}}\!\!\!\!{^{L}}\I^{K}(\e)\\
&&\hspace*{5mm}-\,(V\leftrightarrow -V).\nonumber
\eea
Evaluating this expression and performing the projection, one finds the physical
charge current to be
\bea
j_{L}^{(2)}&=&\frac{\pi}{4}\glr^{2}\left\{3V
-M\left[(V+B)\coth\left(\frac{V+B}{2T}\right)\right.\right.\\
&&\left.\left.\hspace*{24mm}
       -(V-B)\coth\left(\frac{V-B}{2T}\right)\right]\right\}.\nonumber
\eea

The magnetization entering this formula is the {\it on-shell} magnetization
determined in section \ref{magn}, and since we include no dressing of internal
lines this quantity receives no additional renormalization from the retarded and
advanced Dyson equations.

The second order corrections to the correlation function $D_{LR}$, are contained
in the second diagram in Fig.~\ref{current}. After contracting indices with the
measurement vertices, this may be written in terms of Peierls and Cooper bubbles
as
\bea
\lefteqn{D^{K(2)}_{LR}(t,t)=
-\frac{1}{128}J_{L\g}J_{\g R}\,\theta^{P}_{\al\al'\al''}
\!\int\!\!\frac{d\e}{2\pi}}\label{eq:DKLR3}\\
&&\times\left[
\Lambda_{a''\!b}^{c d''}
\ind{\al'\!}{R\,}{{\undl \K}}{ba'}{\!\!d'\!c}(\e)\,
\Lambda_{a\,b''}^{c''\!d}\,
\ind{\al''\!\!}{\g\,\,}{{\undl \K}}{\,b''\!a''}{\!\!\!\!\!\!d''\!c''}(\e)\,
\ind{\al\,}{L}{{\undl \K}}{{\ovrl a'}a}{\!\!\!d\,\,{\ovrl d'}}(\e)\,
\right.\nonumber\\
&&\hspace*{1.5mm}-\left.
\Lambda_{a''\!b}^{c''\!d}\,
\ind{\al'}{L}{{\ovrl {\undl \K}}}{ba'}{\!\!dc'}(\e)\,
\Lambda_{a\,b''}^{c\,d''}\,
\ind{\al''\!\!}{\g\,\,}{{\ovrl {\undl \K}}}{\,b''\!a''}{\!\!\!\!\!\!d''\!c''}(\e)
\,\ind{\al\,}{R}{{\ovrl {\undl \K}}}{{\ovrl a'}a}{\!\!{\ovrl c'}c}(\e)\,
\right],\nonumber
\eea
which upon full contraction leads to
\bea
\lefteqn{D^{K(2)}_{LR}(t,t)=\frac{1}{512 i}J_{L\g}J_{\g R}
\,\theta^{P}_{\al\al'\al''}
\!\int\!\!\frac{d\e}{2\pi}}\\
&&\!\!\times\left[
 {_{\al'\,}}\!\!\!\!\!{^{R}}\I^{K}(\e)
 {_{\al''\!}}\!\!\!\!{^{\g\,}}\I^{A}(\e)
 {_{\al}}\!\!\!\!{^{L}}\I^{A}(\e)
+{_{\al'\,}}\!\!\!\!\!{^{R}}\I^{R}(\e)
 {_{\al''\!}}\!\!\!\!{^{\g\,}}\I^{K}(\e)
 {_{\al}}\!\!\!\!{^{L}}\I^{A}(\e)\right.\nonumber\\
&&\left.\hspace*{36mm}
+\,{_{\al'\,}}\!\!\!\!\!{^{R}}\I^{R}(\e)
 {_{\al''\!}}\!\!\!\!{^{\g\,}}\I^{R}(\e)
 {_{\al}}\!\!\!\!{^{L}}\I^{K}(\e)\right]\nonumber\\
&&\hspace*{10mm}
+\,(V\leftrightarrow -V).\nonumber
\eea
Inserting this into Eq. (\ref{eq:currentdef}), we end up with
\bea
\lefteqn{j^{(3)}_{L}=\frac{\theta^{P}_{\al\al'\al''}}{64i}
J_{LR}^{2}(J_{LL}+J_{RR})\!\int\!\!\frac{d\e}{2\pi}}\\
&&\times
\RE\left[{_{\al}}\!\!\!\!{^{L}}\I(\e)+{_{\al}}\!\!\!\!{^{R}}\I(\e)\right]
\IM\left[{_{\al''}}\!\!\!\!{^{R}}\I(\e)\right]
{_{\al'\!}}\!\!\!\!{^{L}}\I^{K}(\e)\nonumber\\
&&\hspace*{10mm}-\,(V\leftrightarrow -V),\nonumber
\eea
\bwt
\noindent
which may finally be evaluated and projected to obtain the physical current 
\bea
j_{L}&=&\frac{\pi}{4}\glr^{2}
\left\{
      V\left(1+4\gd\ln\frac{D}{|V|} \right)            
+(V+B)\left(1+4\gd\ln\frac{D}{|V+B|}\right)
+(V-B)\left(1+4\gd\ln\frac{D}{|V-B|}\right)\right.\label{eq:Ipert}\\
&&\hspace*{25mm}
-M\left[
\coth\left(\frac{V+B}{2T}\right)
\left\{(V+B)\left(1+2\gd\ln\frac{D}{|V+B|}\right)
      +2V\gd\ln\frac{D}{|V|}
      +2B\gd\ln\frac{D}{|B|}\right\}\right.\nonumber\\
&&\hspace*{30mm}\left.\left.-\coth\left(\frac{V-B}{2T}\right)
\left\{(V-B)\left(1+2\gd\ln\frac{D}{|V-B|}\right)
      +2V\gd\ln\frac{D}{|V|}
      -2B\gd\ln\frac{D}{|B|}\right\}
\,\,\,\,\right]\,\,\right\}\nonumber
\eea
\ewt
Here again, $\ln(D/|x|)$ is a short-hand for $\ln(D/\sqrt{x^2+T^2})$ as all
logarithms are cut off by $T$.
The current has acquired logarithmic corrections, which again enter as
$[1+g\ln(..)]$, and in zero magnetic field we recover the conductance obtained
earlier in Refs.~\onlinecite{Appelbaum66} and~\onlinecite{Kaminski99}
(reinstalling $e$ and $\hbar$):
\be
G(V)=\frac{e^2}{\pi \hbar}(\frac{\pi}{2} \glr)^{2}
     \,3\left(1+4\gd\ln\frac{D}{|eV|}\right). 
\ee

\begin{center}
\begin{figure}[t]
\includegraphics[width=0.95 \linewidth]{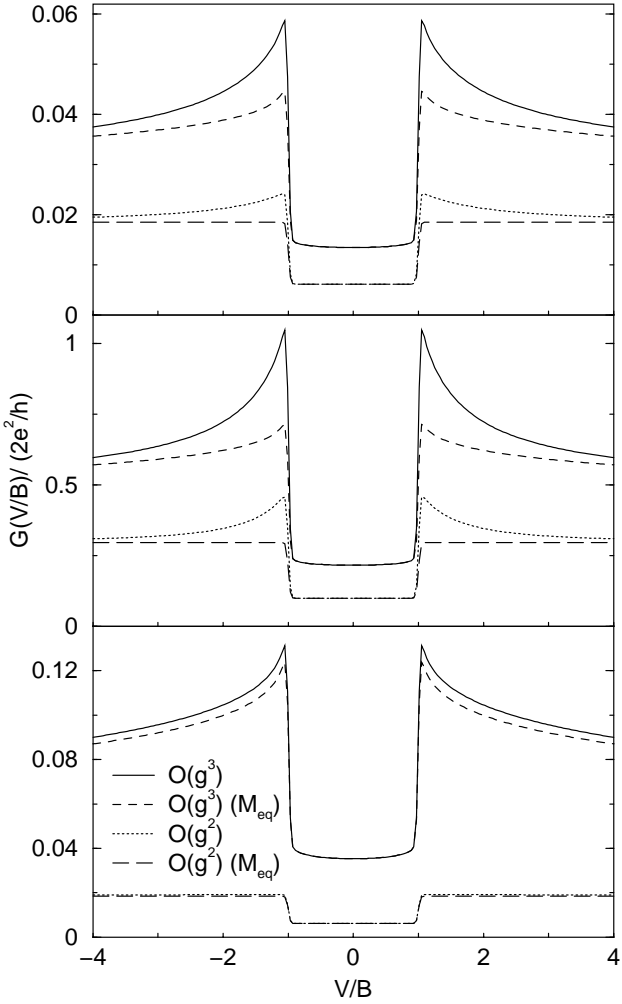}
\caption{\label{condplotsV} Conductance as a function of $V/B$, for $T=10^{-2} B$
and $D=10^3 B$.  Solid and dashed (dotted and long-dashed) lines correspond to
3.(2.) order perturbation theory. Dashed and long-dashed lines have $M=1$ and
correspond to Appelbaum's result, which neglects the $V$-dependence of $M$ shown
in Fig.~\ref{magnplotsV}.
Upper panel: $\gll=\grr=\glr=0.05$.
Middle panel:  $\gll=\grr=0.05$ and $\glr=0.2$.
Lower panel:  $\gll=\grr=0.2$ and $\glr=0.05$.}
\end{figure}
\end{center}

In a finite magnetic field, however, the proliferation of logarithms
is more intricate. While the logarithmic structure in (\ref{eq:Ipert})
was derived already by Appelbaum\cite{Appelbaum66,Appelbaum67a}, he
tacitly employed the equilibrium magnetization $M=\tanh(B/2T)$,
instead of the correct expression (\ref{magAll}), and thereby left out
a number of logarithmic corrections to the resulting conductance.
These early results by Appelbaum are therefore only valid for a local
moment coupled dominantly to one electrode, in which case $g_{LL}\gg
g_{RR},g_{LR}$ and therefore indeed $M=\tanh(B/2T)$, as was pointed
out in section~\ref{magn}. Having solved the quantum Boltzmann
equation in section \ref{magn}, including the leading logarithmic
corrections, we can now correct for this omission. In Fig.
\ref{condplotsV} we plot the conductance as a function of $V/B$, for
$T\ll V$, using both the nonequilibrium magnetization given by
(\ref{magAll}--\ref{eq:MpertN}) and the equilibrium value ($M\approx
1$ since $T\ll B$), corresponding to Appelbaum's result. In both cases
a cusp develops at $|V|\approx |B|$, beyond which spin-flip tunneling
processes are energetically viable. Including the nonequilibrium
magnetization, however, leads to a pronounced enhancement of the
cusps, coming from $\partial M/\partial V$ (shown in
Fig.~\ref{magnplotsV}), which is missing in Appelbaum's result. In the
lower panel, where $\glr <\gd$, Appelbaum's conductance is closer to
the true curve, as for small $\glr$ one is closer to equilibrium
($\glr=0$ is an equilibrium problem).

Our result for the conductance appears to be similar to a plot by Sivan and
Wingreen\cite{Sivan96} for the Anderson model, but since that work does not
contain an explicit analytical expression for the current and the magnetization
it is difficult to compare the two results. Furthermore, logarithmic corrections
to the quantum Boltzmann equation appear not to be included in their approach.

In the high-temperature limit $T\gg\max(V,B)$, there is no effect of
$V$ on the magnetization, and we recover Appelbaum's result for the
current, or rather the result he would have obtained by performing the
final energy integrations in his expressions (39) and (60) of
Ref.~\onlinecite{Appelbaum67a}. The closed form expression presented
in Ref.~\onlinecite{Appelbaum67a} is the conductance rather than the
current, and as an aside we note that differences from his formulas
(where the $T$-dependence arises through hyperbolic tangents) to our
Eq. (\ref{eq:Ipert}) (with hyperbolic {\em cotangents}) arise due to
an (invalid) approximation by Appelbaum, when he approximated the
derivative of the Fermi function by a $\delta$-function. Using the
identity $f(\w+\e)(1-f(\w+\W))=N(\e-\W)(f(\w+\W)-f(\w+\e))$ between
Fermi, and Bose functions, we find $\coth(x)+x \partial_{x}\coth(x)$
instead of $\tanh(x)$ giving rise to a different prefactor ($2/3$
instead of $1$) for $T$ larger than either $V$, $B$ or $|V-B|$.

In the limit of $|V-B|\ll T\ll |V|$, the current may be expanded to give 
\be
j_{L}\approx\frac{\pi}{4}\glr^{2}
B\left( 1+4\gd\ln\frac{D}{2|B|}\right), \ee which remains finite as
$T\to 0$. When expanded in the same limit, the conductance, however,
takes the form \bea \lefteqn{\hspace*{-10mm}
G(B,T)\approx\frac{\pi}{4}\glr^{2}
}\nonumber\\
&&\hspace*{5mm}\times\,\frac{5}{2}\left(1+2 \gd
        \left[\ln\frac{D}{|B|}+\ln\frac{D}{T}-1\right]\right),
\eea
which reveals a logarithmic divergence as $T\to 0$.

This divergence seems to suggest that logarithmic corrections may grow
large, signaling a transition to the strong coupling regime. Such
behavior is, however, prohibited by spin relaxation processes
emanating from the finite current flowing through the dot. As
discussed in Refs.~\onlinecite{Rosch01} and~\onlinecite{Rosch03a}, the relevant
cutoff energy is the voltage dependent transverse spin relaxation rate
$\G_{2}$. At temperatures $T < \G_{2}$ the factor $T$ in $\ln(D/T)$ is
replaced by $\G_{2} \sim g^2V$, leading to a finite correction even in
the limit $T\rightarrow 0$. This cut-off arises as a joint effect of
vertex and self-energy corrections, which will be worked out in detail
in a forthcoming publication\cite{Paaske03b}.

\section{Discussion}\label{discussion}

In the present work we have calculated the local magnetization and
the charge current through a quantum dot at large bias voltage and in
the presence of a magnetic field.  We have considered a Kondo model where
the dot is represented by a quantum spin $S = 1/2$, coupled to leads
by exchange interaction and tunneling.

  Remarkably, the structure of perturbation theory in steady-state
  nonequilibrium is rather different to that in equilibrium. The main
  physical reason is that in our problem the occupation on the dot is
  completely undetermined in the limit of vanishing couplings to the
  leads and therefore has  to be calculated from the solution of the
  quantum Boltzmann equation (i.e. from a self-consistent Dyson
  equation). More generally, in equilibrium all distribution functions
  are given  {\em exactly} by the ``bare'' Boltzmann, Fermi or Bose
  functions without any corrections from interactions while out of
  equilibrium these functions have to be calculated and will depend
  on all the couplings.  Probably the most drastic consequence is that
  for finite voltages the magnetization (\ref{mSec}) is modified even
  for vanishing couplings, i.e., in {\em zeroth} order of perturbation
  theory as has been emphasized in Ref.~\onlinecite{Parcollet02}. As a
  consequence the structure of logarithmic corrections to the
  magnetization is  also rather different out of equilibrium. As
  the matrix elements of order $g^2$ in the quantum Boltzmann equation get
  corrections of order $g^3 \ln(D)$, the perturbative magnetization
  (\ref{magAll}) is generically of the form $\frac{g^2+g^3
    \ln(..)}{g^2+g^3 \ln(..)}\approx {\it const.}+O[g \ln(..)]$, to be
  compared to the equilibrium case where logarithmic corrections arise
  only to order $g^2 \ln(..)$.

In this paper, we have calculated the magnetization and the current through a
Kondo dot to third order in the coupling, including the leading logarithmic
terms. In contrast to earlier treatments of this
problem\cite{Appelbaum66, Appelbaum67a}, the effect of the nonequilibrium
magnetization on the current is incorporated and indeed shown to be important for
a typical quantum-dot experiment in the Kondo regime. In finite magnetic field
the differential conductance exhibits threshold behavior at $V =\pm B$,
reflecting the fact that spin-flip tunneling is possible only for $|V|>|B|$.
Taking the voltage dependence of the magnetization properly into account, the
conductance shows a cusp at $V =\pm B$, already to lowest order in perturbation
theory. Going one order higher, logarithmic corrections are found to enhance
these cusps even further, increasing the conductance substantially over the
threshold plateau, even for magnetic fields much larger than temperature. This
behavior clearly calls for a resummation of the perturbation series to infinite
order.

The intricate structure of logarithmic corrections, revealed by
nonequilibrium perturbation theory, enforces a modified formulation of
the perturbative renormalization group. Thinking in terms of poor man's scaling,
a resummation of the logarithmic corrections cannot simply be achieved by
collecting the logarithmic corrections into a renormalized coupling constant.
This is clearly illustrated by Eq. (\ref{eq:chi}) for the susceptibility, which
shows that different combinations of logarithms appear in the numerator and in
the denominator. However, as we have recently demonstrated~\cite{Rosch03a},
exactly this structure of leading logarithms is generated in a poor man's scaling
approach dealing with energy dependent coupling {\it functions}. A detailed
account of this approach will be given in a subsequent
publication\cite{Rosch03b}.

An additional important difference between nonequilibrium and equilibrium physics
is the fact that  in a nonequilibrium situation the current through the system
generates substantial noise. As a consequence, quantum coherence is limited to
energies above a certain relaxation rate $\G_2$.  The scale $\G_{2}$ will
therefore cut off all logarithmic divergences which remain in the limit $T\to0$
in the perturbative expressions for magnetization and current found in this
paper. This important piece of physics is not
included in the low-order bare perturbation theory presented here and
requires a resummation of subleading self-energy and vertex
corrections. This will be demonstrated explicitly in a forthcoming
publication\cite{Paaske03b}.

Historically, the Kondo effect has played an important role in the
development of techniques like the renormalization
group\cite{Anderson70,Wilson7580} to treat strong-coupling problems
for systems in thermal equilibrium. It is our hope that the present
perturbative calculation of leading logarithmic corrections can serve
both as a starting point and a check for future developments of
similar methods applicable to systems out of equilibrium.

\begin{acknowledgments}

We thank P. Coleman, L. Glazman, C. Hooley, J. Kroha,  and O. Parcollet
for useful discussions. J. P. acknowledges the hospitality of the {\O}rsted
Laboratory at the University of Copenhagen in Denmark, where parts of this work
was carried out. This work was supported in part by the CFN and the Emmy Noether
program (A.R.) of the DFG. 

\end{acknowledgments}

\appendix*

\section{Magnetization and self-energy corrections}\label{renmagn}

 In this appendix we calculate the observable magnetization to order
$g^2\ln(..)$,
by including self-energy corrections to the pseudo fermion spectral function.
Using the prescription (\ref{eq:canonav}) for
evaluation of a canonical ensemble average, the magnetization is determined as
\be
M=\lim_{\la\to\infty}
\frac{n_{\up\lambda}-n_{\down\lambda}}
     {n_{\up\lambda}+n_{\down\lambda}},\label{eq:magnfrac}
\ee
where 
\be
n_{\al\la}=\int^{\infty}_{-\infty}\!\!\frac{d\w}{2\pi}\,
{\cal A}_{\al}(\w)n_{\al\la}(\w),\label{eq:occnumdef2}
\ee
and in principle, the full frequency dependence of both the spectral, and the
distribution function is needed to carry out this integral.

To illustrate how such renormalization works out within the
pseudo fermion approach, we shall commence with the simpler case of
thermal equilibrium and consider merely linear response for $B\ll T$.
Traditionally, most work regarding the Kondo effect on the magnetic
susceptibility in this regime has been conducted using a
Kubo-formula\cite{Yosida65, Brenig70, Okiji70}, and to the best of our
knowledge only one work has taken the approach outlined above to
calculate the magnetization directly from the renormalization of the
$pf$ spectral function\cite{Zawadowski69}. Since, however, this latter
work contains an error, and as we think that the calculation is instructive, we
shall consider this simpler case in some detail before briefly discussing the
nonequilibrium case.

\subsection{Equilibrium Magnetization}

At zero bias-voltage the local spin is in equilibrium with the leads, and the
$pf$ distribution function reduces to a simple Fermi function. In the Keldysh
formalism, this may be viewed as a simple consequence of the KMS
(Kubo-Martin-Schwinger) boundary condition\cite{Rammer86}, which states that in
thermal equilibrium
${\cal G}^{>}_{\al\la}(\w)=-{\cal G}^{<}_{\al\la}(\w)\exp[(\w+\la)/T]$. The
quantum Boltzmann equation (\ref{eq:qbeq}), may be rewritten as
\be
\Sigma^{>}_{\al\la}(\w){\cal G}^{<}_{\al\la}(\w)=
\Sigma^{<}_{\al\la}(\w){\cal G}^{>}_{\al\la}(\w),
\ee
from which the KMS condition is seen to imply that
\be
\frac{\Sigma^{>}_{\al\la}(\w)}{\Sigma^{<}_{\al\la}(\w)}=-e^{(\w+\la)/T}.
\label{eq:kms}
\ee
This can also be verified explicitly from our perturbative expressions for
$\Sigma^{>,<}$, by employing the (KMS) condition
$1-n_{\al\la}=n_{\al\la}\exp[(\la-\al B/2)/T]$ (explicitly satisfied by
a Fermi function) for the on-shell equilibrium occupation numbers. Inserting
(\ref{eq:kms}) into Eq. (\ref{eq:dfctid}) shows that indeed
\be
n_{\al\la}(\w)=\frac{1}{e^{(\w+\la)/T}+1},
\ee
and when derived in this way, it becomes clear that the KMS condition ensures a
highly non-trivial cancellation of interaction corrections in the case of thermal
equilibrium. Applying a finite voltage, this condition is violated and the
nonequilibrium distribution function will be affected by interactions in the
manner which we have described in section~\ref{magn}.

As mentioned earlier, the first order $pf$ Hartree self-energy
vanishes unless one includes a Zeeman term for the conduction
electrons. However, adding such a term,
\be H^{ce}_{\rm  Zeeman}=
B\!\!\!\sum_{{\bf k},\s;{\bf k}',\s';\g}
c^{\dagger}_{\g{\bf k}\s}\tau^{3}_{\s\s'}c_{\g{\bf k}'\s'},\label{eq:cezeem}
\ee to
the Hamiltonian, one finds that the self-energy is entirely real and
given as
\bea
\RE[\Sigma^{(1)}_{\al}(\w)]&=&\frac{1}{4 i}J_{\g\g}
\tau^{i}_{\s\s}\tau^{i}_{\al\al}\int\!\!\frac{d\e}{2\pi}
\Lambda^{cd}_{11}{\undl G}^{dc}_{\g\s}(\e)\nonumber\\
&=&\al B \gd/2,
\eea
where the $ce$ Green-function now depends on the
spin. Note that the Keldysh component of the Hartree self-energy is
identically zero, due to the fact that $\RE\left[G_{\g\al}(\e)\right]$
is an uneven function. In a case where particle-hole symmetry is
broken on an energy-scale $\delta$, one obtains a finite contribution
of order $J\delta/D$, which can be neglected for large $D$.  The only
effect of including a Zeeman term for the conduction electrons is
therefore a constant shift of the $pf$ energy levels.

The second order $pf$ self-energy has a real part, which has only a negligible
contribution from the term (\ref{eq:cezeem}). Without this term, one finds that
\bea
\RE[\Sigma^{(2)}_{\al}(\w)]&=&-\frac{g_{\g\g'}^{2}}{32}\theta_{\al\al'} 
\int_{-D}^{D}\!\!\!d\e\,\tanh\left(\frac{\e+\w-\mu_{\g'}}{2T}\right)\nonumber\\
&&\times
\ln\left(\frac{D^{2}}{(\e-\mu_{\g}-\al' B/2)^{2}+T^{2}}\right),
\eea
which implies that
\be
\RE[\Sigma^{(2)}_{\al}(-\al B/2)]\approx\
\al B g^{2}\ln\frac{D}{T}
\ee
and
\be
\partial_{\w}\RE[\Sigma^{(2)}_{\al}(\w)]\approx
-\frac{3}{2}\,g^{2}\ln\frac{D}{T},
\ee
in the case of $V=0$ and $T\gg B$.

The broadening of the $pf$ energy levels is given by the imaginary part of the
self-energy as $\G=i(\Sigma^{>}-\Sigma^{<})$, and when neglecting $\Sigma^{<}$,
which projects to zero, we end up with $\G=i\Sigma^{>}$ which is given by
\be
\G_{\al}(\w)=g^{2}\theta_{\al\al'}(\w+\al' B/2)(1+N(\w+\al' B/2)),
\ee
to second order in $g$.
This function is highly asymmetric and expanding to first order in $B/T$ one
finds the asymptotic behavior 
\be
\G_{\al}(\w)=\left\{\begin{array}{ll}
3\pi g^{2}\w & ,T\ll\w\\
3\pi g^{2}T\left(1-\frac{\al B}{3T}\right) & ,-T\ll\w\ll T\\
3\pi g^{2}|\w|e^{-|\w|/T}\left(1-\frac{\al B}{6T}\right) &, \w\ll -T
\end{array}\right.
\ee
complemented by the fact that $\G_{\al}(\w)=0$ for $|\w|>2D$, since the
excitation of particle-hole pairs giving rise to the broadening is limited by the
bandwidth. Since $\G$ is essentially constant near the peak of the spectral
function, one may approximate ${\cal A}_{\al}$ in (\ref{eq:renpfA}) by a
Lorentzian for $\w\in[-T,T]$, and by $\G_{\al}(\w)/\w^{2}$ for $|\w|>T$.
Introducing the wave function renormalization factor
\bea
Z_{\al}(\w)&=&\left|1-\partial_{\w}\RE[\Sigma_{\al}(\w)]\right|^{-1}\nonumber\\
&=&1-\frac{3}{2}\,g^{2}\ln\frac{D}{T},
\eea
the coherent part of this approximate spectral function integrates to $Z$ on the
interval $[-T,T]$. The exponential integral of the negative frequency tail
contributes with a number of the order of $g^{2}$, which should be neglected,
while the integral from $T$ to $D$ yields exactly $1-Z$, which ensures
that this approximate spectral function integrates to 1.

To find the magnetization, the integral (\ref{eq:occnumdef2}) may now be
evaluated using the Boltzmann distribution and this approximate spectral
function. The spectral function is centered at a frequency $\w_{\al}$
satisfying the equation $\w_{\al}=-\al B/2+\RE[\Sigma_{\al}(\w_{\al})]$, and
consequently the integral over $[-T,T]$ contributes a factor of
$Z n_{\al\la}(\w_{\al})$, that is
\bea
\lefteqn{\int_{-T}^{T}\frac{d\w}{2\pi}
\frac{\G_{\al}}{(\w-\w_{\al})^{2}+(\G_{\al}/2)^{2}}Z n_{\al\la}(\w)
}\nonumber\\
&\approx &Z+\al\frac{B}{2T}\left(1-g-\frac{7}{2}\,g^{2}\ln\frac{D}{T}\right),
\label{eq:Cint}
\eea
to first order in $B/T$. Since the distribution function falls off exponentially
for $\w\gg T$, the integral over $[T,D]$ is negligible. On the negative frequency
tail, the spectral function decays exponentially, but this is compensated in the
integral by the exponential increase of the Boltzmann distribution, and one finds
that
\bea
\lefteqn{\int_{-D}^{-T}\frac{d\w}{2\pi}
\frac{\G_{\al}}{(\w)^{2}}n_{\al\la}(\w)}\nonumber\\
&\approx &1-Z-\al\frac{B}{4T}g^{2}\ln\frac{D}{T}.\label{eq:Lint}
\eea
Adding up (\ref{eq:Cint}) and (\ref{eq:Lint}) and inserting in
(\ref{eq:magnfrac}), one ends up with the magnetization
\be
M=\frac{B}{2T}\left(1-g-4\,g^{2}\ln\frac{D}{T}\right),\label{eq:finalMeq}
\ee
to which one may finally add the corresponding induced spin-polarization of the
conduction electrons to obtain the total magnetization
\bea
\lefteqn{\hspace{-1.6cm}\langle S^{z}+s^{z}_{L}+s^{z}_{R}\rangle-
\langle s^{z}_{L}+s^{z}_{R}\rangle_{0}^{\rm Pauli}}\nonumber\\
&\approx &
\frac{B}{4T}\left(1-2g-4\,g^{2}\ln\frac{D}{T}\right).
\eea
Here the $z$-component of the total spin of lead $\g$ has been introduced as
\be
s^{z}_{\g}=\!\!\!\sum_{{\bf k},\s;{\bf k}',\s';\g}\!\!\!
c^{\dagger}_{\g{\bf k}\s}\tau^{3}_{\s\s'}c_{\g{\bf k}'\s'},
\ee
and the $pf$ and $ce$ g-factors are assumed to be equal. This result matches the
high-temperature expansion of the exact Bethe ansatz solution as it should
\cite{Andrei83, Hewson93}.

In Ref.~\onlinecite{Zawadowski69}, the normalization of the approximate spectral
function is demonstrated just like here. In calculating the magnetization,
however, the contribution (\ref{eq:Lint}) was not included and the $Z$-factor was
argued to be canceled by the same factor appearing in the denominator of
$(\ref{eq:magnfrac})$. Altogether, this error leads to a prefactor of 2 instead
of 4 in front of the $g^{2}\ln(D/T)$-term in (\ref{eq:finalMeq}), which destroys
the correspondence with the exact result.


\subsection{Nonequilibrium Magnetization}

As demonstrated in the equilibrium case, the magnetization is renormalized in
a delicate balance between shifts and broadening of the $pf$ energy levels, i.e.
between the influence of $\RE[\Sigma]$ on the coherent part, and of $\G$ on the
incoherent tails of the spectral function. In the case of finite bias-voltage,
however, the renormalization of the distribution function becomes important.

At finite voltage, there are logarithmic corrections to $\Sigma^{<}$
and $\Sigma^{>}$, which no longer cancel in Eq. (\ref{eq:dfctid}), and
the magnetization now exhibits the stronger renormalization by factors
of $g\ln(D)$ rather than $g$ and $g^{2}\ln(D)$. To properly describe
the crossover to equilibrium, as $T$ becomes greater than $V$, one
should include these subleading corrections deriving from the
renormalization of the $pf$ spectral function. This can be done in
much the same way as above, as long as care is taken to separate the
coherent part of the spectral function from the incoherent tails at
either $T$ or $V$, so as to ensure normalization of the total spectral
weight.  When $T$ is increased beyond $V$, the $g\ln(D)$ corrections
get less and less important.

It should be emphasized that, as mentioned already in
section~\ref{tuncurrent}, the effects of shifts and broadening,
discussed in this appendix, affect only the observable magnetization
and have no influence on our result for the leading logarithmic
corrections to the current.



\begin{thebibliography}{99}

\bibitem{Hewson93}A. C. Hewson, {\it The Kondo Problem to Heavy Fermions}
(Cambridge University Press, 1993).

\bibitem{Kondo64}J. Kondo,
Prog. Theor. Phys. {\bf 32}, 37 (1964).

\bibitem{Andrei8082}N. Andrei,
Phys. Rev. Lett. {\bf 45}, 379 (1980);
Phys. Lett. {\bf 87A}, 299 (1982).

\bibitem{Wilson7580}K. G. Wilson,
Rev. Mod. Phys. {\bf 47}, 773 (1975);
H. R. Krishna-murthy, J. W. Wilkins and K. G. Wilson,  
Phys. Rev. {\bf B21}, 1003 (1980).

\bibitem{Anderson70}P. W. Anderson,
J. Phys. C {\bf 3}, 2436 (1970).

\bibitem{Glazman88}L. Glazman and M. Raikh,
JETP Letters {\bf 47}, 452 (1988).

\bibitem{Ng88} T. Ng and P.A. Lett,
Phys. Rev. Lett. {\bf 61}, 1768 (1988).

\bibitem{Goldhaber98}D. Goldhaber-Gordon, Hadas Shtrikman, D. Mahalu,
David Abusch-Magder, U. Meirav and M. A. Kastner,
Nature {\bf 391}, 156 (1998).

\bibitem{Cronenwett98} S. M. Cronenwett, T. H. Oosterkamp and L. P. Kouwenhoven,
Science {\bf 281}, 540 (1998).

\bibitem{Schmid98} J. Schmid, J. Weis, K. Eberl and K. von Klitzing,
Physica {\bf B258}, 182 (1998).

\bibitem{Nygaard00}J. Nyg{\aa}rd, D. H. Cobden and P. E. Lindelof,
Nature {\bf 408}, 342 (2000).

\bibitem{vanderWiel00}W. G. van der Wiel, S. De Franceschi, T. Fujisawa,
J. M. Elzerman, S. Tarucha, and L. P. Kouwenhoven,
Science {\bf 289}, 2105 (2000).

\bibitem{Goldhaber98b}D. Goldhaber-Gordon, J. G\"{o}res, M. A. Kastner,
Hadas Shtrikman, D. Mahalu and U. Meirav,
Phys. Rev. Lett. {\bf 81}, 5225 (1998).

\bibitem{Shen68}L. Y. L. Shen and J. M. Rowell,
Phys. Rev. {\bf 165}, 566 (1968).

\bibitem{Nielsen70}P. Nielsen,
Phys. Rev. B {\bf 2}, 3819 (1970).

\bibitem{Appelbaum72}J. Appelbaum and L. Y. Shen
Phys. Rev. B {\bf 5}, 544 (1972).

\bibitem{Wallis74}R. H. Wallis and A. F. G. Wyatt,
J. Phys. C: Solid State Phys. {\bf 7}, 1293 (1974).

\bibitem{Bermon78}S. Bermon, D. E. Paraskevopoulos and P. M. Tedrow,
Phys. Rev. B {\bf 17}, 210 (1978).

\bibitem{Magno77}R. Magno and J. G. Adler,
Phys. Rev. B {\bf 15}, 1744 (1977).

\bibitem{Wolf70}E. L. Wolf and D. L. Losee,
Phys. Rev. B {\bf 2}, 3660 (1970).

\bibitem{Wolf75}E. L. Wolf, in {\it Solid State Physics}, eds. H. Ehrenreich,
F. Seitz and D. Turnbull (Academic, New York, 1975), vol. 30, p.1.

\bibitem{Appelbaum66}J. Appelbaum,
Phys. Rev. Lett. {\bf 17}, 91 (1966).

\bibitem{Appelbaum67a}J. Appelbaum,
Phys. Rev. {\bf 154}, 633 (1967).

\bibitem{Anderson66}P. W. Anderson,
Phys. Rev. Lett. {\bf 17}, 95 (1966).

\bibitem{Losee69}D. L. Losee and E. L. Wolf,
Phys. Rev. Lett. {\bf 23}, 1457 (1969).

\bibitem{Ivezic80}T. Ivezi\'{c},
J. Magn. Magn. Mater. {\bf 15-18}, 933 (1980).

\bibitem{Solyom68}J. S\'{o}lyom and A. Zawadowski,
Phys. Kondens. Materie {\bf 7}, 325 (1968);
Phys. Kondens. Materie {\bf 7}, 342 (1968).

\bibitem{Appelbaum70}J. Appelbaum and W. F. Brinkman
Phys. Rev. B {\bf 2}, 907 (1970).

\bibitem{Zawadowski67}A. Zawadowski,
Phys. Rev. {\bf 163}, 341 (1967).

\bibitem{Appelbaum69}J. Appelbaum and W. F. Brinkman
Phys. Rev. {\bf 186}, 464 (1969).

\bibitem{Ivezic75}T. Ivezi\'{c},
J. Phys. C: Solid State Phys., {\bf 8}, 3371 (1975).

\bibitem{Caroli71}C.Caroli, R. Combescot, P. Nozieres and D. Saint-James,
J. Phys. C: Solid St. Phys., {\bf 4}, 916 (1971).

\bibitem{Wolf85}E. L. Wolf, {\it Principles of Electron Tunneling Spectroscopy},
(Oxford University Press, Oxford, 1985).

\bibitem{Meir93}Y. Meir, N. S. Wingreen and P. A. Lee,
Phys. Rev. Lett. {\bf 70}, 2601 (1993).

\bibitem{Wingreen94}N.S. Wingreen and Y. Meir,
Phys. Rev. B {\bf 49}, 11040 (1994).

\bibitem{Hettler94}M. H. Hettler, J. Kroha and S. Hershfield,
Phys. Rev. Lett. {\bf 73}, 1967 (1994);
Phys. Rev. B {\bf 58}, 5649 (1998).  

\bibitem{Norlander99}P. Nordlander, M. Pustilnik, Y. Meir, N. S. Wingreen and
D. C. Langreth,
Phys. Rev. Lett. {\bf 83}, 808 (1999);

\bibitem{Plihal00}M. Plihal, D. C. Langreth, P. Nordlander,
Phys. Rev. B {\bf 61}, R13341 (2000);

\bibitem{Krawiec02}M. Krawiec and K. I. Wysokinski
Phys. Rev. B 66, 165408 (2002)

\bibitem{Sivan96}N. Sivan and N. S. Wingreen,
Phys. Rev. B {\bf 54}, 11622 (1996).

\bibitem{Goldin98}Y. Goldin and Y. Avishai,
Phys. Rev. Lett. {\bf 81}, 5394 (1998); Phys. Rev. B {\bf 61}, 16750 (2000).

\bibitem{Kaminski99}A. Kaminski, Yu. V. Nazarov and L. I. Glazman,
Phys. Rev. Lett {\bf 83}, 384 (1999); Phys. Rev. B {\bf 62}, 8154 (2000).

\bibitem{Schiller95}A. Schiller and S. Hershfield,
Phys. Rev. B {\bf 51}, 12896 (1995); {\it ibid}. {\bf 58}, 14978 (1998);
K. Majumdar, A. Schiller and S. Hershfield,
Phys. Rev. B {\bf 57}, 2991 (1998).

\bibitem{Emery92}V. J. Emery and S. Kivelson,
Phys. Rev. B {\bf 46}, 10812 (1992).

\bibitem{Konik01}R. M. Konik, H. Saleur and A. Ludwig,
Phys. Rev. Lett. 87, 236801 (2001); Phys. Rev. B 66, 125304 (2002).

\bibitem{Rosch03a}A. Rosch, J. Paaske, J. Kroha and P. W\"olfle, 
Phys. Rev. Lett. {\bf 90}, 076804 (2003)

\bibitem{Appelbaum67b}J. Appelbaum, J. C. Phillips and G. Tzouras,
Phys. Rev. {\bf 160}, 554 (1967).

\bibitem{Nagaoka65}Y. Nagaoka,
Phys. Rev. {\bf 138}, A1112 (1965).

\bibitem{Abrikosov65}A. A. Abrikosov,
Physics {\bf 2}, 5 (1965), {\it ibid}. {\bf 2}, 61 (1965) 

\bibitem{Suhl65}H. Suhl,
Phys. Rev. {\bf 138}, A515 (1965);
H. Suhl and D. Wong,
Physics {\bf 3}, 17 (1967).

\bibitem{Rosch03b}A. Rosch, J. Paaske, J. Kroha and P. W\"olfle, 
in preparation.

\bibitem{Zawadowski69}A. Zawadowski and P. Fazekas,
Z. Physik {\bf 226}, 235 (1969).

\bibitem{Rammer86}J. Rammer and H. Smith,
Rev. Mod. Phys. {\bf 58}, 323 (1986).

\bibitem{Kadanoff62}L. P. Kadanoff and G. Baym,
{\it Quantum Statisticsl Mechanics} (Benjamin, New York, 1962).

\bibitem{Langreth76}D. C. Langreth in
{\it Linear and Nonlinear Electron Transport in Solids},
eds. J. T. Devreese and E. Van Doren (Plenum, New York, 1976). 

\bibitem{Parcollet02}O. Parcollet and C. Hooley, 
Phys. Rev. B {\bf 66}, 085315 (2002).

\bibitem{Rosch01}A. Rosch, J. Kroha and P. W\"{o}lfle,
Phys. Rev. Lett. {\bf 87}, 156802 (2001).

\bibitem{Paaske03b}J. Paaske, A. Rosch and P. W\"{o}lfle,
in preparation.

\bibitem{Yosida65}K. Yosida and A. Okiji
Prog. Theoret. Phys. {\bf 34}, 505 (1965).

\bibitem{Brenig70}W. Brenig, J. A. Gonzalez, W. G\"{o}tze and P. W\"{o}lfle,
Z. Physik {\bf 235}, 52 (1970).

\bibitem{Okiji70}A. Okiji, A. Kato and H. Shiba,
Suppl. Prog. Theoret. Phys.{\bf 46}, 182 (1970).

\bibitem{Andrei83}N. Andrei, K. Furuya and J. H. Lowenstein,
Rev. Mod. Phys. {\bf 55}, 331 (1983).

\end{thebibliography}
\end{document}